\documentclass[10pt,journal,compsoc]{IEEEtran}

\ifCLASSOPTIONcompsoc
  \usepackage[nocompress]{cite}
\else
  \usepackage{cite}
\fi
\ifCLASSINFOpdf
  \usepackage[pdftex]{graphicx}
  \graphicspath{{./figures/}{./faces/}}
  \DeclareGraphicsExtensions{.pdf,.jpeg,.png}
\else
\fi
\usepackage{amsmath}
\usepackage{url}

\newcommand\MYhyperrefoptions{bookmarks=true,bookmarksnumbered=true,
pdfpagemode={UseOutlines},plainpages=false,pdfpagelabels=true,
colorlinks=true,linkcolor={black},citecolor={black},urlcolor={black},
pdftitle={Data-driven sparse skin stimulation can convey social touch information to humans},
pdfsubject={Haptics},
pdfauthor={M Salvato et al.},
pdfkeywords={social communication, data-driven actuation, tactile devices, social distancing}}
\usepackage[\MYhyperrefoptions,pdftex]{hyperref}
\usepackage{algorithm}
\usepackage{algpseudocode}
\makeatletter
\algrenewcommand\ALG@beginalgorithmic{\footnotesize}
\makeatother
\begin{document}
%
\title{Data-driven sparse skin stimulation can convey social touch information to humans}
%
%
%
%

\author{M. Salvato, Sophia R. Williams, Cara M. Nunez, Xin Zhu, Ali Israr, Frances Lau, Keith Klumb,\\ Freddy Abnousi, Allison M. Okamura, Heather Culbertson
\IEEEcompsocitemizethanks{\IEEEcompsocthanksitem M. Salvato (corresponding author: msalvato@cs.stanford.edu; \url{web.stanford.edu/\~msalvato}) and A. M. Okamura are with the Department of Mechanical Engineering, Stanford University, Stanford, CA 94305.\protect
\IEEEcompsocthanksitem S. R. Williams is with the Department of Electrical Engineering, Stanford University, Stanford, CA 94305.\protect
\IEEEcompsocthanksitem C. M. Nunez is with the Departments of Bioengineering and Mechanical Engineering, Stanford University, Stanford, CA 94305.\protect
\IEEEcompsocthanksitem X. Zhu and H. Culbertson are with the Department of Computer Science, University of Southern California, Los Angeles, CA 90089\protect
\IEEEcompsocthanksitem A. Israr, F. Lau, K. Klumb, and F. Abnousi are with the Facebook Inc., 1 Hacker Way, Menlo Park, CA 94025\protect

}}
\IEEEtitleabstractindextext{%
\begin{abstract}
During social interactions, people use auditory, visual, and haptic cues to convey their thoughts, emotions, and intentions. Due to weight, energy, and other hardware constraints, it is difficult to create devices that completely capture the complexity of human touch. Here we explore whether a sparse representation of human touch is sufficient to convey social touch signals. To test this we collected a dataset of social touch interactions using a soft wearable pressure sensor array, developed an algorithm to map recorded data to an array of actuators, then applied our algorithm to create signals that drive an array of normal indentation actuators placed on the arm. Using this wearable, low-resolution, low-force device, we find that users are able to distinguish the intended social meaning, and compare performance to results based on direct human touch. As online communication becomes more prevalent, such systems to convey haptic signals could allow for improved distant socializing and empathetic remote human-human interaction.
\end{abstract}

\begin{IEEEkeywords}
social communication, data-driven actuation, tactile devices, social distancing
\end{IEEEkeywords}}

\maketitle

\IEEEdisplaynontitleabstractindextext

%
\IEEEpeerreviewmaketitle

\ifCLASSOPTIONcompsoc
\IEEEraisesectionheading{\section{Introduction}\label{sec:introduction}}
\else
\section{Introduction}
\label{sec:introduction}
\fi

\IEEEPARstart{S}{ocial} touch is a natural mode of communication between humans, and its importance is becoming more evident as we increase remote communication through videoconference, email, and text messages. Without the depth and subtlety achievable in person, remote communication causes increased feelings of loneliness and social isolation~\cite{mcpherson2006social}, which can have a significant impact on mental and physical health~\cite{holt2015loneliness}. The need to social distance~\cite{Kisslereabb5793} exacerbates this issue, and further highlights the importance of more expressive remote communication. 

Previously, researchers have sought ways to virtually replicate the feelings of social touch through wearable and holdable haptic devices. Several methods of conveying touch have been explored, such as vibrations~\cite{seifi2013first}, thermal stimulation~\cite{wilson2016hot}, and air jets~\cite{tsalamlal2015haptic}. However, most social touch haptic devices have focused on replicating a single form of social touch, such as a handshake~\cite{nakanishi2014remote}, a hug~\cite{mueller2005hug,tsetserukou2010haptihug}, or a stroke on the arm~\cite{tsalamlal2015haptic,eichhorn2008stroking,CulbertsonHS2018,knoop2015tickler,israr2018towards}. In addition, most of these devices leverage manually generated signals, often optimized for simplicity of description (such as frequency and intensity), rather than conveying the richness and subtlety of touch interaction. Humans, on the other hand, have shown moderate success interpreting touch cues from partners even without context~\cite{hauser2019uncovering, hertenstein2009communication}, and their social touch signals can be complex and varied. Even devices that directly map human interaction from a sensor to an actuator have relied on tight coupling between the sensing and actuation methods~\cite{nakanishi2014remote, eichhorn2008stroking, rantala2013touch}. Rather than focusing on specialized hardware, our method seeks to richly represent a range of social touch messages by using a data-driven method to leverage recorded human social touch data. This allows for complex and varied social touch signals using a variety of recording and actuation devices.

Social haptic devices vary in the type of contact they achieve with the skin. Some devices are limited to applying haptic stimulation, such as vibration, heat, or pressure, to a local region of the skin~\cite{seifi2013first,wilson2016hot}. On the other end of the spectrum are those which can move continuously along the skin, applying stimulation anywhere over a range of skin~\cite{tsalamlal2015haptic,eichhorn2008stroking}. We focus on what we define as ``sparse'' devices, which apply haptic stimulation multiple, separate, local regions of the skin~\cite{huisman2013tasst, CulbertsonHS2018, NunezTOH2019, israr2018towards}. We focus on sparse devices because they can be less mechanically complex than continuous devices, but more expressive than applying stimulus to only a single local skin region.

This paper explores whether social touch data can be represented using sparse actuation given appropriate processing, allowing for the creation of systems that can represent a variety of social touch messages mapped directly from recorded human interactions. We develop a consistent and generalizable, data-driven algorithm to map recorded human touch to a sparse representation. We leverage multi-object tracking techniques~\cite{Zhang2008GlobalDA} to find data likely to represent sustained or high-pressure touch interactions, then find optimal regions to render on a haptic system. We display the sparse touch signal with eight fixed points of contact, via an arm-worn array of one-degree-of-freedom voice coil actuators, each 37-50~mm apart. These contact points are at a larger distance than the threshold for discrimination by afferents in the skin, based on direction discrimination tests~\cite{ackerley2014touch}. We tested participants' ability to recognize social touch scenarios based on passively felt signals produced by the actuator array.

This work was performed as a 2-stage study. In the first stage (Section 2), we created a novel social touch dataset. This dataset involved human-human interaction with a provided scenario, gave no instruction on which gestures the human should use, and recorded force data directly. In related studies, human social touch interaction was recorded with defined gestures on a mannequin~\cite{Jung2014TouchingTV} and small robot~\cite{Flagg2013AffectiveTG}. These data involved simple gestures, with explicit instruction to the user of how each gesture was to be performed. They were collected with the primary goal of classification~\cite{jung2015touch}, rather than generative modeling. Hauser et al.~\cite{hauser2019uncovering} used camera-based recording and electromagnetic trackers to measure touch direction and movement associated with six thoughts to be conveyed for open-ended interaction. However, the study lacked force sensing, and used only single-word prompts. We provide scenarios so participants can touch in an informed way with similar context across pairs.

In the second stage of our study, we applied a mapping algorithm to instances from our social touch dataset (Section 3) and examined if a new set of participants could use the sensation they felt to determine the scenario the data was drawn from (Section 4). We found that participants achieved 45\% accuracy on 6 scenarios, in comparison with 57\% accuracy found for human-human interaction between close friends and partners~\cite{hauser2019uncovering}. This suggests that social touch signals can be represented via sparse skin stimulation.

The contributions of this work are as follows. The social touch classification accuracy in our human subject test was achieved via the use of a novel social touch dataset that was passed through a data-driven mapping algorithm and rendered on a sparse haptic device. We describe a detailed account of our data collection, and  provide the resulting dataset for public use. We also present a novel rendering algorithm and provide the associated code to facilitate future work in haptic rendering. Because our patterns were found to be interpretable, we posit it is possible that sparse touch signals are sufficient for humans to classify social touch.
\vspace{-.2\baselineskip}
\begin{figure}[t]
  \includegraphics[width=\columnwidth]{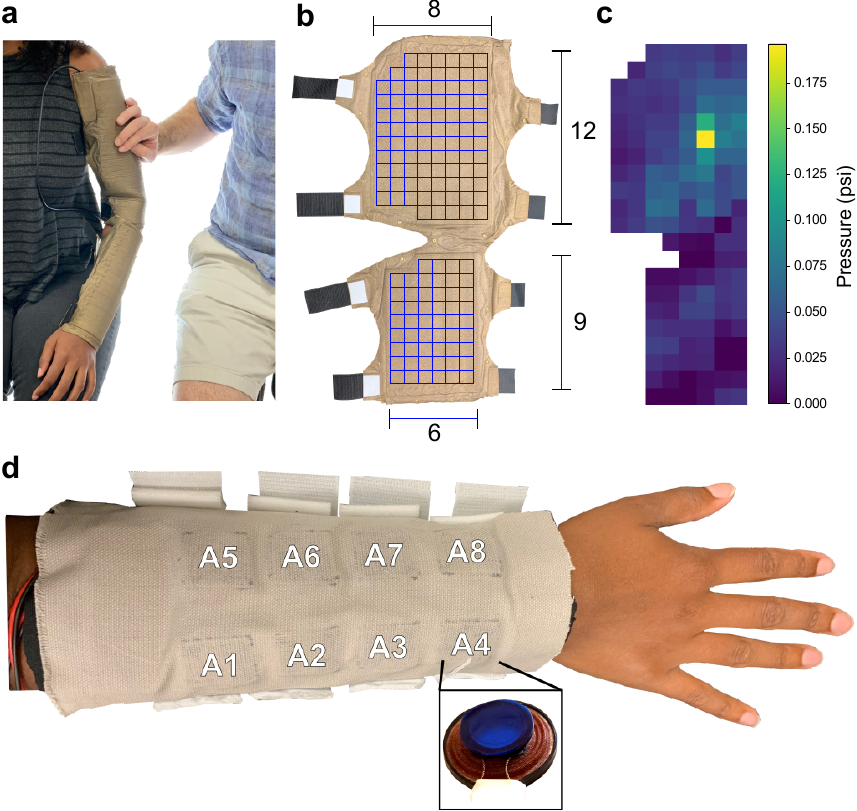}
  \vspace{-\baselineskip}
  \caption{Data recording and actuator hardware. (\textbf{a-c}) Social touch data recording setup and sensor. (a) Image of the experimental setup. One participant wore a pressure sensor sleeve while the other participant touched the sensor sleeve arm to express social meaning. Complementary social scenarios were played over headphones to participants to cue the social touch. (b) Image of a sensor sleeve. The sensor sleeve is shown laid flat, with squares indicating the individual sensor cells. Each cell is 1~$\text{in}^\text{2}$. This is the larger of two sleeves used during the study, to account for differing arm sizes. The smaller sleeve has upper arm dimension 8x10 and lower arm dimension 6x8. (c) An image of a single data frame from the sensor. The sensor can record pressure as high as 2.96 psi at 20~Hz. (\textbf{d}) Actuator sleeve. An image of a user wearing our actuator sleeve, with an example actuator. The actuator is covered with a thin piece of rigid plastic. While the sensor covers the upper arm and forearm, the actuators are limited to the forearm.}
  \label{fig:hardware}

\end{figure}
\section{A Naturalistic Social Touch Dataset}

To analyze human social touch and create artificial touch signals, we collected a naturalistic social touch dataset. Participants were recruited in pairs who self-identified as either close friends or romantic partners  who felt comfortable interacting with each other through touch. We recruited participants through Stanford University email lists that included hundreds of students and postdocs enrolled in various programs of study. Participants gave informed consent, and the protocol was approved by the Stanford University Institutional Review Board (protocol \#22514). Twenty pairs of partners were recruited, for 40 participants total (17 male, 23 female, ages 22-34 years old). Thirty-five participants self-reported that they were right-handed, three participants that they were left-handed, and two participants that they were ambidextrous. Eleven of the pairs were close friends. Out of these, three pairs were male-female, three were male-male, and five were female-female. Nine of the pairs were romantic partners. Out of these, eight pairs were male-female and one pair was female-female. No male-male romantic partners were tested. All pairs included at least one participant who was affiliated with Stanford.

\begin{figure}
\centering
  \includegraphics[width=.8\columnwidth]{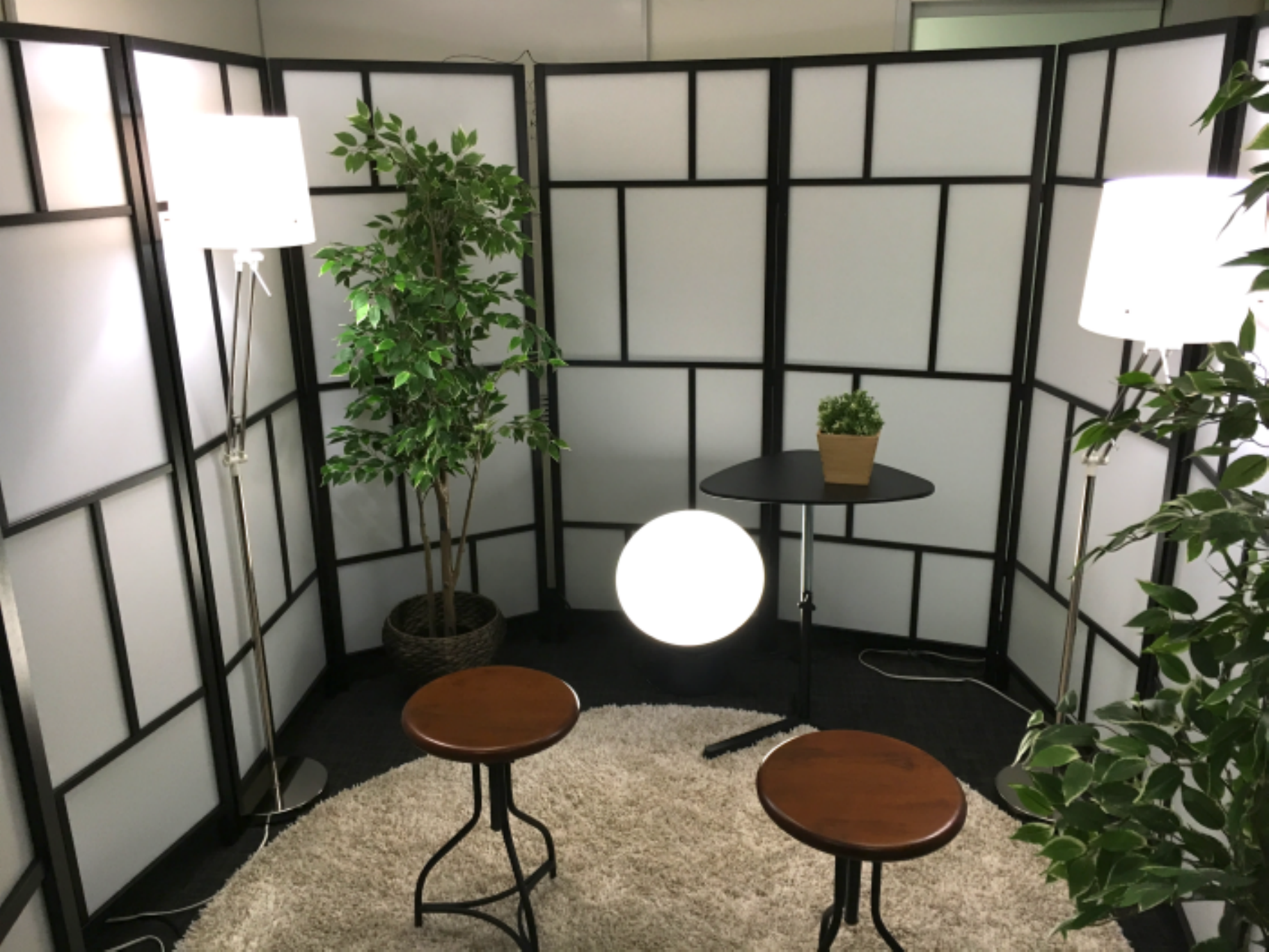}
  \vspace{-\baselineskip}
  \caption{Image of social touch recording room. The participants sat on the two provided stools. All lights in the room were color changing bulbs, which changed hue based on the prompt.
  \vspace{-.5\baselineskip}}
  \label{fig:room}
  \vspace{-\baselineskip}
\end{figure}

\begin{table}
\centering
\caption{Scenario identifiers and associated touch meaning categories. Our 6 scenarios span 5 meaning categories. For the repeated category (``Announcing a response'') we have a positive and a negative valence example. These terms motivated our scenarios and were presented to the toucher and receiver prior to the scenario.}
\vspace{-3mm}
\begin{tabular}{lll}
Scenario Identifier & Touch Meaning Category\\
\hline
    Attention seeking & Attention-getting\\
    Gratitude & Appreciation\\
    Happiness & Announcing a response (positive valence)\\
    Calming & Support\\
    Love & Affection\\
    Sadness & Announcing a response (negative valence)\\
\hline
\end{tabular}

\label{fig:touch_categories}
\vspace{-1\baselineskip}
\end{table}

A dataset entry involves one participant wearing a soft pressure sensor on their arm, and another touching them on the worn sensor to express a response to a described scenario (Fig.~\ref{fig:hardware}a). We used detailed scenarios in an attempt to keep the participants maximally invested, and so all pairs imagined similar contexts. We attempted to create a setting where participants were comfortable expressing themselves by creating a room where subjects were isolated from the experimenters, artificial plants and a rug were placed in the room, and floor and lamp lighting were provided (Fig.~\ref{fig:room}). We focused on the interaction from one person's hand to another's arm, based on prior work of socially acceptable places to give and receive social touch \cite{suvilehto2015topography} and informal pilot testing. We placed no restriction on the duration or nature of their movements beyond interaction on the arm to prevent the subjects from feeling constrained in their actions. With this population we expect that participants were able to map the provided scenario to similar ones they previously experienced with that partner. This data collection scenario is unique compared to other datasets~\cite{Jung2014TouchingTV,Flagg2013AffectiveTG}, which involve a human interacting with a non-living object. While \cite{huisman2013tasst} similarly uses recorded touch on a human-worn sensor, they were performed by the experimenter in a controlled setting and limited to simple gestures. They refer to this as a controlled "best case" recording scenario.

\begin{figure*}
\centering
  \includegraphics[width=\textwidth]{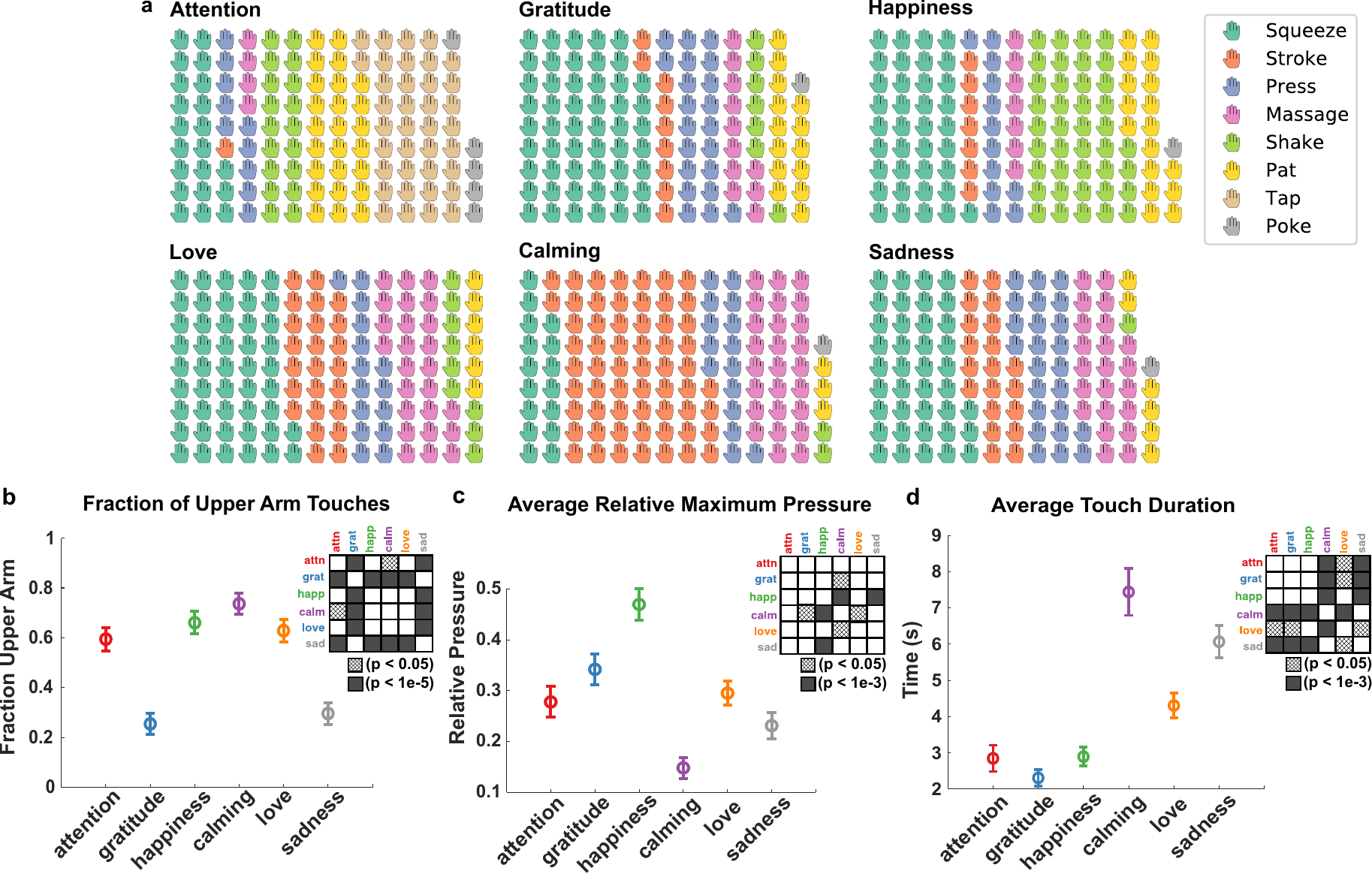}
  \vspace{-\baselineskip}
  \caption{Dataset gesture counts and per scenario metrics. (\textbf{a}) Gesture counts. For each touch, we manually labelled the gestures apparent in the touch. This plot shows how many times each gesture appeared for each touch. Patterns appear in the data, such as a large number of taps and pats for attention, strokes for calming, and shakes for happiness. (\textbf{b}) Touch location plot. The fraction of touches for which more frames have maximum value on upper arm as opposed to lower arm. Using Friedman test we find the relationship between scenario and upper arm count was significant ($\chi^2(5, N=666) = 98.06$, $p = 1.4\textrm{e-}19$). The associated significance matrix shows the probability two scenarios differ and was calculated using pairwise Friedman tests. Note that Friedman is equivalent to Cochren Q and McNema tests for the overall and pairwise data respectively~\cite{spss16}. (\textbf{c}) Relative maximum pressure plot. For each participant we determined the maximum pressure for each of their 18 touch instances. We then normalized these values to obtain a relative maximum pressure for each instance for each participant. Using Friedman test we find the relationship between scenario and relative maximum pressure was significant ($\chi^2(5, N=666) = 83.01$, $p = 2.0\textrm{e-}16$). The associated significance matrix shows the post-hoc pairwise probability two scenarios differ and was calculated using the Dunn-Sidak test. (\textbf{d}) Touch duration plot. Participants were free to take as long as desired when touching.  Using Friedman test we find the relationship between scenario and touch duration was significant ($\chi^2(5, N=666) = 285.81$, $p = 1.1\textrm{e-}59$). The associated significance matrix shows the post-hoc pairwise probability two scenarios differ and was calculated using the Dunn-Sidak test. The error bars on (b), (c), and (d) show standard error. We selected our statistical tests as we found the data as non-normal, not homoscedastic, and does not have independence (because each subject responds to every condition)}
  \label{fig:dataset}
  \vspace{-\baselineskip}
\end{figure*}

\begin{figure*}[t!]
  \includegraphics[width=\textwidth]{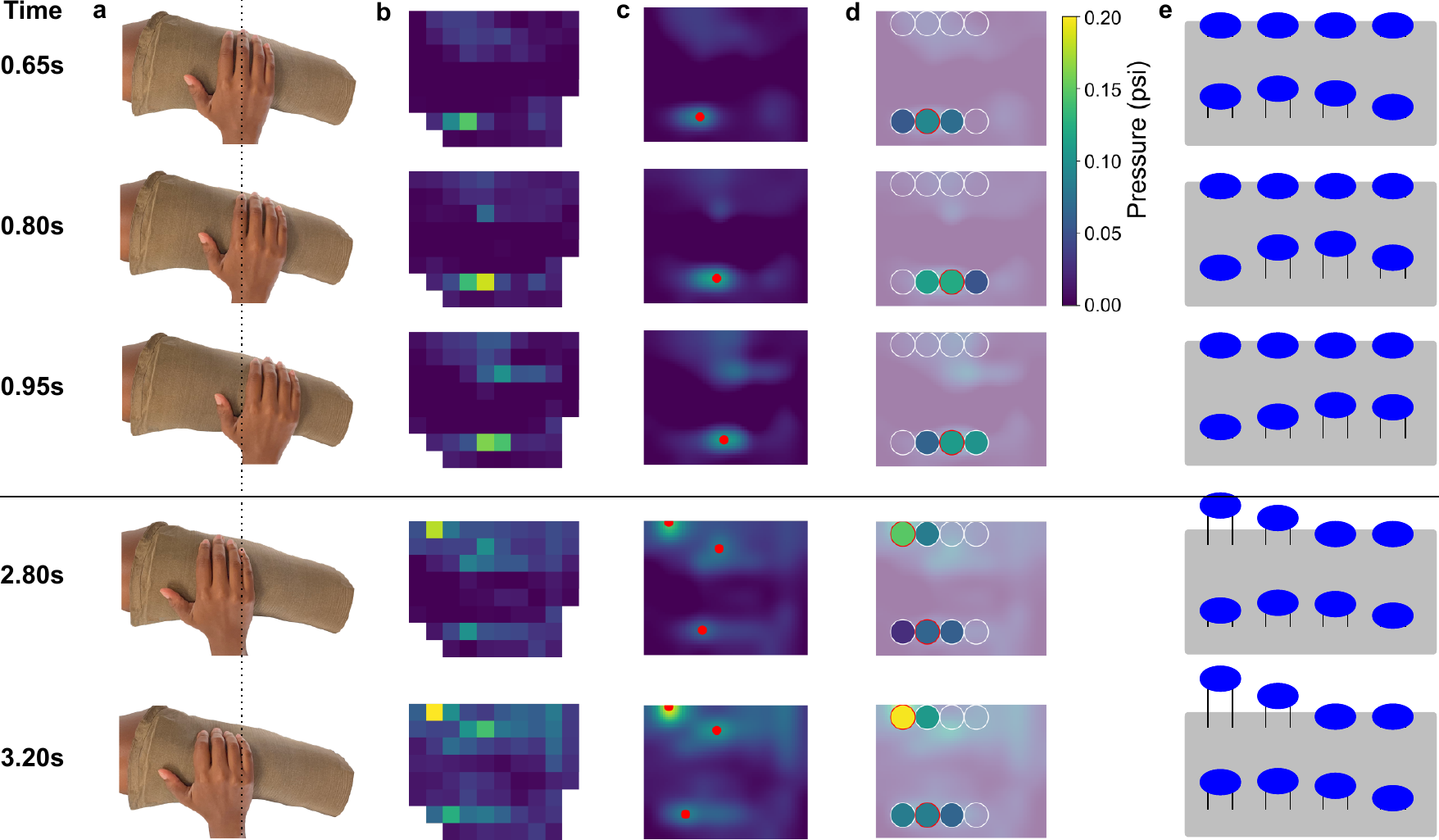}
  \vspace{-\baselineskip}
  \caption{Data mapping algorithm. Here we show representative key frames from our data mapping algorithm. The actual data was processed at 20~Hz. (\textbf{a}) A user interacting with the pressure sensor sleeve. The top three frames show a sampling of a stroking motion, while the bottom two frames show a sampling after the user transitions to a squeezing motion. (\textbf{b}) Example data frames from the sensor sleeve. This is sensor data from a `love' example in our dataset -- it is not drawn from (a) itself. (\textbf{c}) The data from (b), after preprocessing and multi-object tracking. The red dots represent tracked paths. This is the result of Step 1 of our algorithm. (\textbf{d}) Actuator restricted tracking. The circles represent the optimal areas to render, given the tracked paths. The fill color represents the intensity to render. The red circle contains the trajectory and renders its intensity. This is the result of Step 2 of our algorithm. We also render the maximum intensity within the circles which neighbor the trajectory for continuity.  The pressure scale for (b-d) is shown to the right of column (d).  The algorithm is run only once on the data and generates both the stroke and squeeze motions.  (\textbf{e}) Example voice coil activity. This is a representation of our $2\times4$ voice coil sleeve, with each blue circle being a normal indentation actuator. The coils have height proportional to the intensities of each workspace in (d).}
  \label{fig:algorithm}
  \vspace{-\baselineskip}
\end{figure*}

We based our set of scenarios on previous work studying direct social touch between individuals.~\cite{jones1985naturalistic} found that social touch can be divided by meaning into 12 categories, such as support touches, appreciation touches, and affection touches. To have a variety of scenarios, we leveraged prior work in emotion communication through touch~\cite{hertenstein2009communication,guest2011development}, and selected 6 scenarios spanning 5 of the categories presented by~\cite{jones1985naturalistic} (Table~\ref{fig:touch_categories}). For the repeated category we have a positive valence and a negative valence example. Our final set of scenarios is: attention seeking, gratitude, happiness, calming, love, and sadness.

For each pair of participants, one participant, the ``receiver'', wore a soft pressure sensor wrapped on their arm. For each scenario, the lighting level and color changed slightly to match the scenario (Fig.~\ref{fig:room}). The receiver and the other participant, the ``toucher'', each listened to an audio prompt. The receiver listened to a prompt that prepared them to be touched, and the toucher listened to a different prompt describing the scenario and which ended by telling them to touch the receiver. The toucher touched the receiver on the sensor array using one hand. The toucher was told to touch for as long as they wished, and to return their hands to their lap when finished (Supplementary Mov.~SM1). This procedure was repeated for each scenario presented in pseudorandom order in three sets, for a total of 18 interactions. The process was then repeated with the toucher and receiver roles switched. While we refer to scenarios by single word identifiers, our results apply to the \textit{scenarios} presented to the participants. Text and audio recordings of the prompts are available at Supplementary Information: Social Touch Dataset and Supplementary Dataset SD1 respectively. As an example, below is the \textit{love scenario} presented to the toucher:
\vspace{.2\baselineskip}
\begin{quote}
    \noindent \textit{Love. Imagine this: You and the person sitting next to you are spending the afternoon together. You're walking to get a bite, the weather is amazing, and you're catching up on everything in the way that friends do. You look at them, and it suddenly strikes you how much this friendship means to you, that life is so much easier and better with them around. Reach out and touch this person to express your love for them.}
\end{quote}
\vspace{.2\baselineskip}

Our protocol was designed in collaboration with IDEO (Palo Alto, CA, US https://www.ideo.com/). Behavioral scientists at IDEO developed storyboards of the scenarios with the goal of creating an immersive experience and eliciting authentic touches. Various scenarios were iterated in preliminary tests to be natural and compelling. Through these iterations, a number of design choices were made, including: manually advancing scenarios by the experimenter when the users returned their hands to their laps; placing temporary walls  around the participants so they felt alone in the room, comfortable and fully immersed; and telling participants that they could make facial expressions, but not communicate verbally.

The touch data were recorded using a custom soft capacitive sensor sleeve created by Pressure Profile Systems (Los Angeles, CA, US). The sensor wraps around the upper arm and forearm, and records pressure data in 1~$\textrm{in}^2$ cells at 20~Hz. The device has a range from 0psi to 2.96psi  with 0.004 $\pm$ 0.001psi measured resolution. It has calibrated linearity of 99.9 $\pm$ 0.1\% and signal-to-noise ratio of 732 $\pm$ 219. We found when the sensor was laid flat with no contact, pressure readings did not exceed $\pm$0.08psi. We used two sensor sizes, with 120 or 142 sensor cells (Fig.~\ref{fig:hardware}b). The sensors had different lengths and circumferences, with the sensor chosen based on fit for each participant's arm. Fig.~\ref{fig:hardware}c shows a single example data frame from the sensor. We also recorded the interaction from two different views using RGB cameras.

Due to recording errors, the data for three participants was lost, resulting in 37 participants' worth of recorded data. A total of 661 instances were recorded (5 were removed due to additional recording error). Each touch can be described as a series of short gestures (such as ``poke'' or ``stroke''). The experimenters visually inspected the video and time-synchronized pressure sensor waveforms to annotate the recorded data with gestures based on the gesture descriptions in~\cite{hertenstein2009communication}, using a subset of gestures from that work. Fig.~\ref{fig:dataset}a shows the set of gestures with total number of times each gesture was seen across all instances of each scenario, and Fig.~\ref{fig:dataset}b-d shows statistics about touch location, pressure, and duration. To test the importance of the relationship between the toucher and receiver on touch motion, we checked the significance of relationship type (``friends'' or `''romantic partners'') for each of these statistics. We found that touch duration was significantly long for those who identified as romantic partners using the Kruskal-Wallis test ($\chi^2(1, N=222) = 37.23$, $p = 1.0\textrm{e-}9$). Those in romantic relationships may also touch the lower arm more often ($\chi^2(1, N=222) = 3.76$, $p = 5.2\textrm{e-}2$).  We chose the Kruskal-Wallis test as responses were independent, but the data was non-normal and not homoscedastic.

Before the experiment, each participant completed a survey about their experience and preferences in expressing social touch (Supplementary Fig.~S1). Survey data was obtained for all 40 participants. Raw data of all sensor recordings and gesture annotations is provided in a free online dataset (Supplementary Dataset SD2) to be leveraged for future research into both naturalistic remote human-human communication and human-robot interaction.

\section{Data to Actuator Mapping}
We developed an algorithm to map recorded touch data to signals that can be rendered on a haptic device (Fig.~\ref{fig:algorithm}). This algorithm assumes the haptic device consists of an array of actuators to render touch information, and the number of actuators is smaller than the number of sensors. Thus, we reduce the sensor data to the dimensionality of the actuators by finding consistent, high pressure trajectories through the data, and render a selection of those trajectories with the actuator array. For this project we used this algorithm to map to a $2\times4$ array of voice coil actuators (Fig.~\ref{fig:hardware}d, Sec.~\ref{sec:actuator}). Our algorithm is akin to a multi-object tracking problem (Step 1), with actuator specific restrictions (Step 2). 

Step 1 is to find the most salient, contiguous signal within the data. This step is independent of the haptic device and produces a set of signal trajectories through the data. We base this on a multi-object tracking algorithm~\cite{Zhang2008GlobalDA}. This step relies on two assumptions. First, we assume that for a set of data frames the intensity of a point increases monotonically with the probability of interaction with the device. This assumption allows us to capture trajectories with higher intensity. Second, we assume that when points close in time are of similar intensity, they are more likely to represent a sustained interaction with the device. The multi-object tracking algorithm works by solving an optimization problem for which longer, greater intensity paths are weighted more highly. Fig.~\ref{fig:algorithm}c shows a sampling of frames from a tracked path. Fig.~\ref{fig:algorithm}b shows corresponding frames sampled from the raw data.

Step 2 is to find the optimal set of trajectories to render on the haptic device. This step requires each actuator to have a defined workspace, making it dependent on the actuator hardware being used. Workspaces are specified as 3D geometric shapes, for which we require that only one trajectory pass through a given workspace at a given time. For our actuators these are cylinders, which we represent as circles in Fig.~\ref{fig:algorithm}d. To optimize the rendering, we determine a placement of the actuator workspaces within the trajectory space. We determine the placement by testing different translations of the workspaces in the trajectory space, optimized by a scoring function based on the probability of elements in a trajectory and their locations in the workspace. Depending on the actuators and desired use case, the workspaces could be tested under other transformations, such as rotation, scaling, or separation. Fig.~\ref{fig:algorithm}d shows the trajectories tracked from Fig.~\ref{fig:algorithm}c, optimally mapped to the workspaces for our haptic device. Fig.~\ref{fig:algorithm}e shows an example set of actuators rendering those trajectories. We see the algorithm captured both a stroking motion (top) and squeezing motion (bottom).

We could have merged Step 1 and Step 2 into a single optimization where the trajectories are selected with information about the actuator workspaces provided. However, we separated these procedures so that Step 1 represents the trajectories strictly as a function of the data, and therefore is a more accurate representation of the data itself. We also render on actuators that neighbor the actuators for which the trajectory is specified, and perform other processing for smoothness. Below we provide pseudocode and further explanation. Supplementary Information: Mapping Algorithm Formalization provides a full algorithm formalization and Supplementary Dataset SC1 provides the code.

\begin{algorithm*}
\caption{Pressure Data to Actuators Trajectories}
\begin{algorithmic}[1]

\Procedure{MappingAlgorithm}{F, A, M}\\
\Comment{Pressure Data Frames $F$, Actuator Workspaces $A$, Actuator Transforms $M$}
\State $\mathcal{G} \gets \{\} $
\ForAll{$f_t \in F$} \Comment{Get local maxima points across frames.}
\State $\mathcal{G} \gets \mathcal{G} \cup FindMaxima(f_t)$
\EndFor
\State $\mathcal{J}, \mathcal{R}  = MultiObjTracking(\mathcal{G}, F)$ \Comment{Find trajectory set $\mathcal{J}$; determine associated scores $\mathcal{R}$ based on trajectory element intensities.}
\ForAll{$m \in M$}  \Comment{For each transform, find best trajectories and associated scores}
\State $\mathcal{J}_m, \mathcal{R}_m \gets ConstrainedTrajs(\mathcal{J}, \mathcal{R}, m(A))$ \Comment{when restricted to one trajectory per actuator per timestep.}
\EndFor 
\State $\mathcal{J}^*\gets \mathcal{J}_i\text{ s.t. }\underset{R_i \in \mathcal{R}_i}{\sum} R_i \geq \underset{R_j \in \mathcal{R}_j}{\sum} R_j, i \neq j$ \Comment{Select best scoring restricted transformed trajectories.}
\EndProcedure \Comment{In case $\underset{R_i \in \mathcal{R}_i}{\sum} R_i = \underset{R_j \in \mathcal{R}_j}{\sum} R_j$, $\mathcal{J}^*$ chosen arbitrarily.}
\end{algorithmic}
\label{alg:mot}
\end{algorithm*}

\vspace{-.5\baselineskip}
\subsection{Actuation Hardware}
\label{sec:actuator}

Our actuator sleeve consists of eight voice coils (Tectonic Elements TEAX19C01-8), arranged in a $2\times4$ pattern (Fig.~\ref{fig:hardware}d, Supplementary Mov.~SM2). Coils were placed approximately 37~mm apart in each row, and 50~mm apart in each column. The sleeve has an inelastic fabric backing, with softer elastic fabric covering the backs of the voice coils. The sleeve was wrapped around the forearm and affixed using velcro strips, so the voice coils sat on the dorsal side of the arm. The coils were controlled with a PCI board (SENSORAY Model 826) at 1000~Hz, through a linear current amplifier with a gain of 1~A/V via an LM675T op-amp.
\vspace{-.5\baselineskip}
\subsection{Algorithm Pseudocode}

Algorithm~\ref{alg:mot} shows a pseudocode of the mapping algorithm. The algorithm takes 3 inputs: a series of pressure frames $F$, a list of actuator workspaces $A$, and a list of actuator transforms $M$. Each frame $f_t \in F$ consists of 2D coordinates and 1D pressure identical in structure to a single channel image where the pixel value is the pressure. We will refer to these coordinates as pixels. These frames do not need to be rectangular. Each $a_i \in A$ is a 3D shape representing the workspace of actuator $i$. In our work the workspaces are cylinders (seen as circles in 2D). The actuator transforms $m \in M$ are functions which inform how we can place the actuator workspaces relative to the pressure frames. They can consist of operations such as translation, rotation, or scaling, but we only use translation in this work.

\begin{figure*}[t!]
  \includegraphics[width=\textwidth]{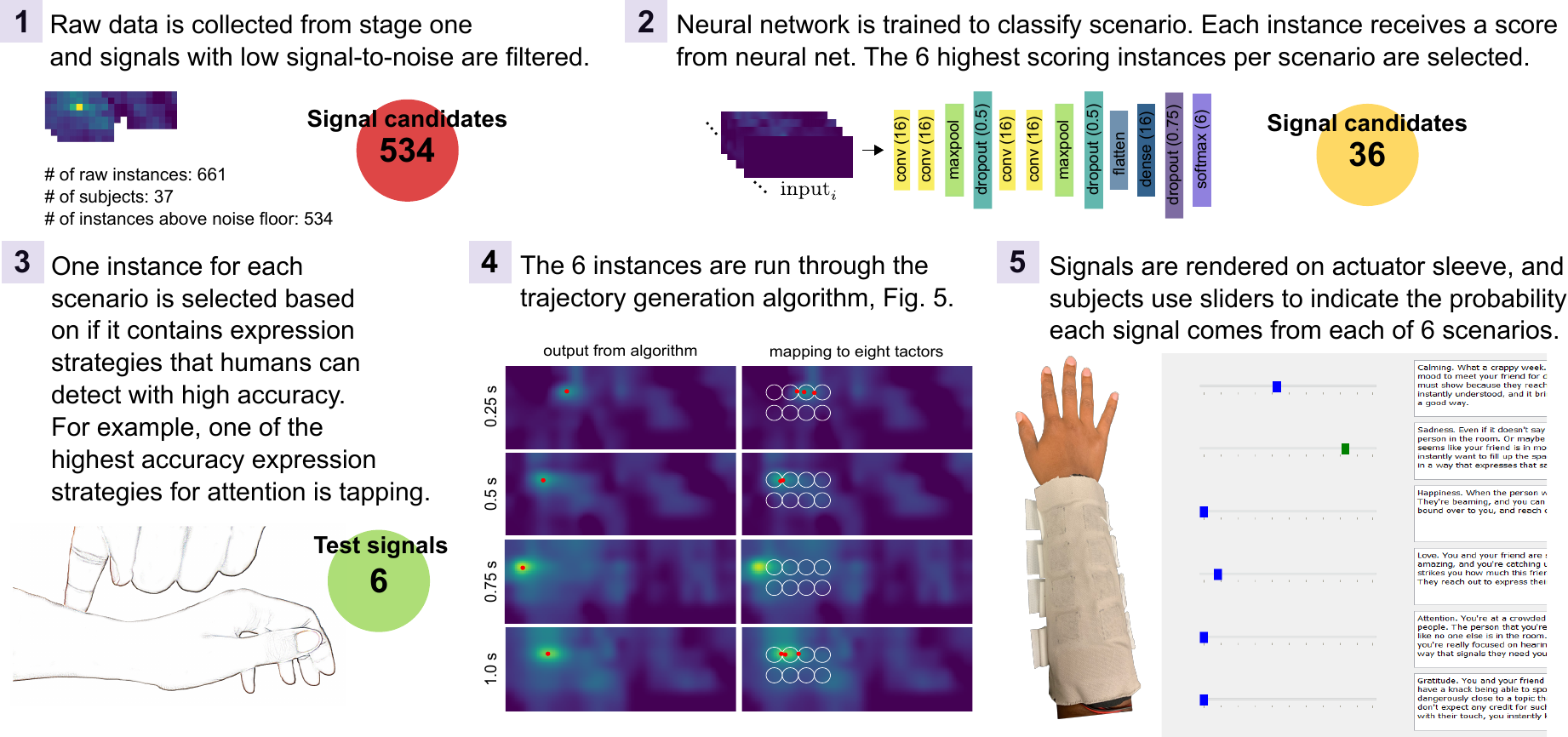}
  \vspace{-\baselineskip}
  \caption{Process to integrate data, mapping algorithm, and hardware for scenario classification experiment. In step 1, we removed sequences for which 5 or fewer total readings exceed .08 psi.}
  \label{fig:workflow}
  \vspace{-\baselineskip}
\end{figure*}

Lines 3-5 the algorithm find the set of all local maxima coordinates $\mathcal{G}$ for the data frames $F$. These are the pixels which have higher pressure than each of their spatial neighbors for the frame in which they are located. This is so the following step is computationally tractable.

Line 6 addresses the multi-object tracking algorithm portion. This is Step 1 described above. Multi-object tracking algorithms are used in computer vision to maintain a consistent labeling of identified objects in a scene across multiple frames~\cite{Zhang2008GlobalDA}. These assume a mechanism already exists to identify the objects. For example, in video of multiple people walking in a busy street, a detector may be used to identify people, then multi-object tracking could be used to track each individual across frames. In our work we seek to find ``trajectories,'' which are sequences of pixels through time that represent sustained, high pressure data. We assume these represent the movement of fingers or the palm on the sensor. As an input, we take the local maxima coordinates $\mathcal{G}$ and data frames $F$. We output salient trajectories $\mathcal{J}$ in the data, which we use to represent the movement of the fingers or palm. Each trajectory also has an associated score $R$, with the set of scores denoted $\mathcal{R}$ for associated trajectories $\mathcal{J}$. We create a set of trajectories $\mathcal{J}$ that are tracked through the data frames $f_t \in F$. A single trajectory $J \in \mathcal{J}$ consists of a sequence of pixels, $p_1, ..., p_l$, through frames $f_{s}, ..., f_{s+l-1}$. $p_i \in f_{i + s-1}$. $l$ represents the length of the trajectory, and $s$ the time of the first frame in the trajectory. A trajectory consists of only contiguous frames, and a single trajectory cannot consist of more than one pixel in a frame. Additionally the pixels are only those in $\mathcal{G}$ -- those which were local maxima from lines 3-5. 

In order to find the trajectories in line 6, we adapt a multi-object tracking algorithm~\cite{Zhang2008GlobalDA} with the following assumptions. First, we assume that for a set of data frames the intensity of a point increases monotonically with the probability of interaction with the device. This assumption allows us to capture trajectories with higher frame intensity. Second, we assume that when points close in time are of similar intensity, they are more likely to represent a sustained interaction with the device. The multi-object tracking algorithm works by solving an optimization problem for which longer, higher intensity trajectories are weighted more highly. The scores $R \in \mathcal{R}$ are also based on the length and intensity of the trajectories (see Supplementary Information: Algorithm Formalization for exact calculation).

Lines 7-8 address the actuator workspace restriction, the first part of Step 2 of our algorithm. This step finds the best sets of trajectories for which each actuator is constrained to act on only one trajectory at a time. For each call inside the loop, we take as input the trajectories $\mathcal{J}$, scores $\mathcal{R}$, and the actuator workspaces. For each iteration of the loop, a single transform $m()$ is applied to all workspaces in $A$. Considering the trajectories as tracking movement in Euclidean space, the actuator workspaces are placed in this space such that the tracked trajectories describe movements for the actuators. Each transform $m()$ then represents a transformation, such as translation, of the set of actuators in the space, so the actuators will render different parts of the trajectories based on the transform. The algorithm then calculates the best subset of the trajectories for that transform, $\mathcal{J}_m$, for which no two trajectories exist in the same workspace at the same time, using the sum of trajectory scores. The trajectory scores are affected by a small weighting based on the actuator workspaces, as explained in Supplementary Information: Algorithm Formalization.

Line 9 then completes Step 2 by selecting the trajectories output by the transform on line 8 which resulted in the highest summed trajectory score. We take this opportunity to highlight why it is important to test different actuator transforms. As an example, consider four consecutive workspaces, and one trajectory long enough to pass through all of them. If the trajectory starts in the first actuator, it would render through all four. However, if the trajectory started in the second actuator, the last segment of the trajectory would not be rendered, since it would be placed in the space beyond the actuators. This problem becomes more difficult with multiple trajectories and the restriction of one trajectory per actuator per timestep, necessitating the development of Step 2 of our algorithm.

\subsection{Data Processing}

For our system, to preprocess the recorded data, we first upsample the data by a factor of seven. We then perform a 2D Gaussian blur with standard deviation of three.

After the algorithm determines which actuators render the trajectory points at each time, we also examine the row-wise neighboring actuators. Each neighboring actuator which is not already rendering a trajectory will render the maximum sensor intensity within their workspace area. This is to effectively smooth the signal, which creates continuous sensations for gestures such as stroking for our system.

After producing the the signals for each actuator to render, we remove sharp peaks in the data. We do this by finding triplets of points that rise and drop by $\frac{1}{8}$ the total pressure range between each pair of frames and setting them to the mean of the first and third point. We then filter the whole signal with a 4~Hz, 5th order Butterworth lowpass filter. During informal pilot testing prior to implementing these measures, participants found it difficult to parse the signals due to jitter, and we saw limits on the speed of the voice coils. This filtering removes the jitter while still maintaining signal information.

\section{Scenario Classification Experiment}

\subsection{Methods}

We conducted an experiment to determine whether participants could classify haptic signals generated by applying our algorithm to our dataset, felt through the voice coil sleeve hardware (Fig.~\ref{fig:workflow}). Six signals were chosen and used in the experiment (Sec.~\ref{sec:selection}), one for each of the six scenarios: attention seeking, gratitude, happiness, calming, love, and sadness. For each signal subjects selected which scenario it was derived from. Participants gave informed consent, and the protocol was approved by the Stanford University Institutional Review Board (protocol \#22514). 30 participants engaged in the experiment (14 female, 16 male, ages 19-35 years old). Participants in the social touch dataset collection were excluded from this experiment.

\begin{figure*}
  \includegraphics[width=\textwidth]{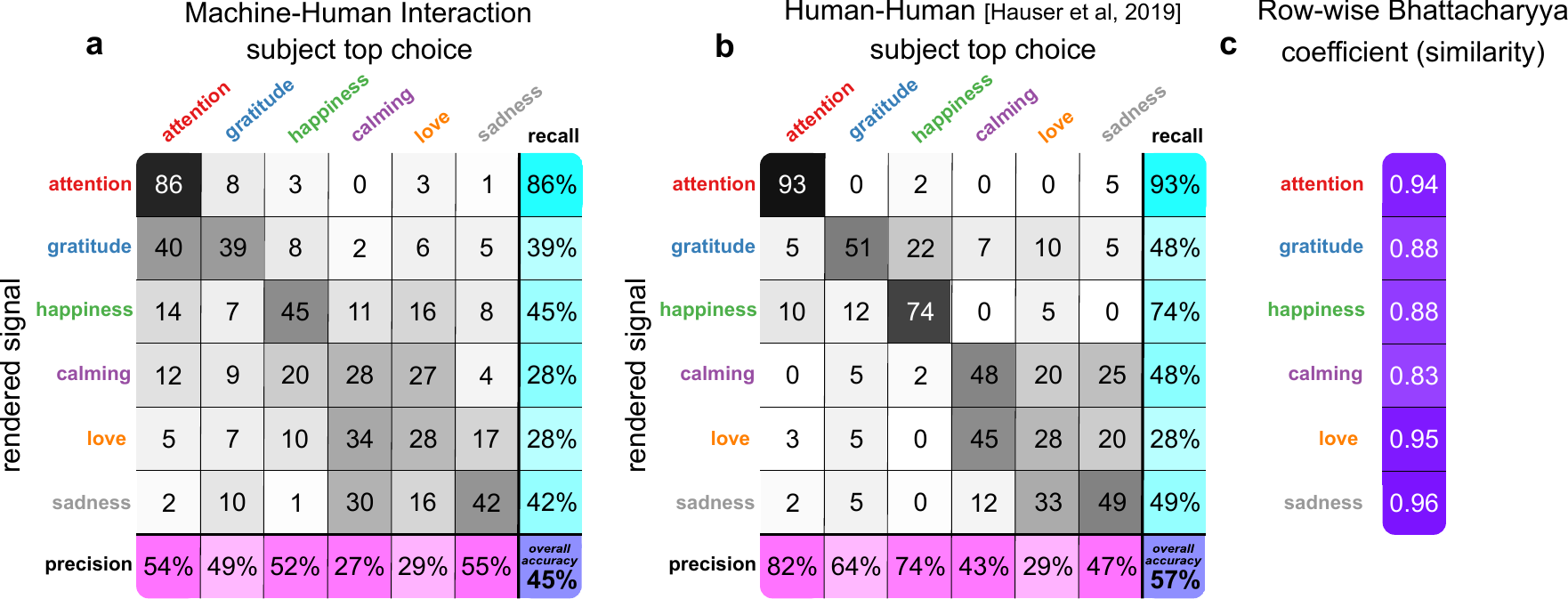}
  \vspace{-\baselineskip}
  \caption{Scenario classification top choice results. (\textbf{a}) The confusion matrix for the scenario that participants thought was most likely displayed by the rendered signal. This provides information on the participants' overall accuracy in identifying the scenario the rendered signal was drawn from, how participants confused signals, and participants' precision and recall for their top scenario choices. Recall ($\frac{\textrm{True Positive Rate}}{\textrm{True Postive Rate} + \textrm{False Negative Rate}}$) gives a sense of how often the rendered signal type was correctly identified when presented, and precision ($\frac{\textrm{True Positive Rate}}{\textrm{True Postive Rate} + \textrm{False Positive Rate}}$) gives a sense of how often that signal type was actually presented when it was chosen. Rows are normalized to 100, with 120 samples per row in raw data. We fit a general linear mixed effects (GLME) model with a binomial distribution and a logit link function where scenario is represented as a 6-level factor and subject as a random effect. The Analysis of Deviance (Type II Wald $\chi^2$ test) shows that the scenario has a significant effect on accuracy ($\chi^2(5) = 87.7$, Pr$(> \chi^2) < 2.2e-16$), implying that scenario affected response. We checked linearity and normality of residuals by inspecting Pearson residuals and the q-q plot respectively. On a per scenario basis, binomial tests show that subjects did not choose randomly (at a significance level of $p < 0.002$) for each scenario. (\textbf{b}) Human-human interaction results from \cite{hauser2019uncovering} for comparison. (\textbf{c}) Row-wise Bhattacharyya coefficients. For each rendered signal row we normalize the values to 1 and then calculate the Bhattacharrya coefficient~\cite{bhattacharyya1946} between the machine-human and human-human row values. This provides comparison of the similarity of responses per scenario. Higher value indicates more similarity, coefficient has range of [0,1].}
  \label{fig:top1}
  \vspace{-\baselineskip}
\end{figure*}

During the experiment, participants listened to the toucher audio for each of the six scenarios. They then passively felt each of the six haptic signals once, with no indication of what the signals represented, so participants were acclimated to the range of signals they would feel. They were then presented each signal a total of three times, in randomized sets of each of the six signals. For each signal, participants were asked to provide the probability that they believed the sensation was drawn from each of the six scenarios using a slider bar. Participants were informed that the values would be normalized to one, and were required to not have an exact tie for the largest value. During this phase of the experiment, subjects were shown text transcriptions of the scenarios to serve as a reminder of the audio. While they were provided the scenario identifier term, they were asked to focus on the scenario itself. In the last phase of the experiment, participants felt each signal once more and were asked to rate the valence and arousal of each signal, using the Self-Assessment Manikins~\cite{bradley1994measuring}. Raw data of participant's responses is provided for free online (Supplementary Dataset SD3). The toucher audio prompts were presented to the participants instead of the receiver audio prompts, because the participant's task was to determine the intent of toucher (device), which is not necessarily encoded in the receiver audio prompts. 

To test the number of subjects required for a moderate effect size significant relationship between provided scenario and response, we used a $\chi^2$ test with 25 degrees of freedom. To get moderate effect size of $w = 0.3$ with power $1-\beta = 0.95$ and error probability $\alpha = 0.05$, we would require 370 samples. We collected approximately double this number of (720) scenario classifications.

\subsection{Instance selection}
\label{sec:selection}

For this experiment we selected instances from the social touch dataset to render on our hardware that exemplified common gestures, as determined by our gesture annotations (Fig.~\ref{fig:dataset}a) and previous work~\cite{hauser2019uncovering}. Because our dataset had over 500 sequences, we used a 3D convolutional neural network (Fig.~\ref{fig:workflow}.2) to classify the instances by scenario, in order to prune the dataset for instances that exemplified the scenario and make instance selection more tractable. We ran 5-fold cross validation to get a classification score for each instance for each scenario. Each sequence was scored based on the fold where it was part of the test set.

The longest sequence we input is 705 data frames. For all sequences shorter than this, we prepended frames with all values set to 0, so the sequence length becomes 705. To address the difference in size of the two pressure sensors, for the shorter sleeve we repeated the column immediately prior to the elbow joint and two columns immediately after the elbow joint. This prevented disruption of the movement along each arm segment. Additionally, because the sensor arrays are not rectangular, each frame was padded with 0s to form a rectangular shape. Therefore, the input data to the neural net was of dimension $(705, 8, 21)$ (Fig.~\ref{fig:hardware}b).

Each convolutional layer was 3D and had filter size $(1,2,2)$ with stride length $(1,1,1)$ and a ReLU activation. Each max pooling layer had pool size $(2,2,2)$ and stride length $(2,2,2)$. Categorical cross entropy loss was used. The Adam optimizer \cite{kingma2014adam} with learning rate $0.001$ was used (along with all other Keras defaults) for 30 epochs.

We found that the neural net attributed high classification scores to touch instances with the gestures we expected, so for each scenario we chose one instance from the top six highest scoring instances for that scenario. These instances were selected based the experimenters' intuition about which signals would be most interpretable by the subjects. After selecting one expression instance per scenario, we processed each with our data mapping algorithm. We chose a time range to render for each signal, but otherwise the processing was identical -- no parameters varied between signals. This selection process resulted in a set of six total signals, one per scenario (Supplementary Figs.~S6-S7).

\begin{figure}
  \includegraphics[width=\columnwidth]{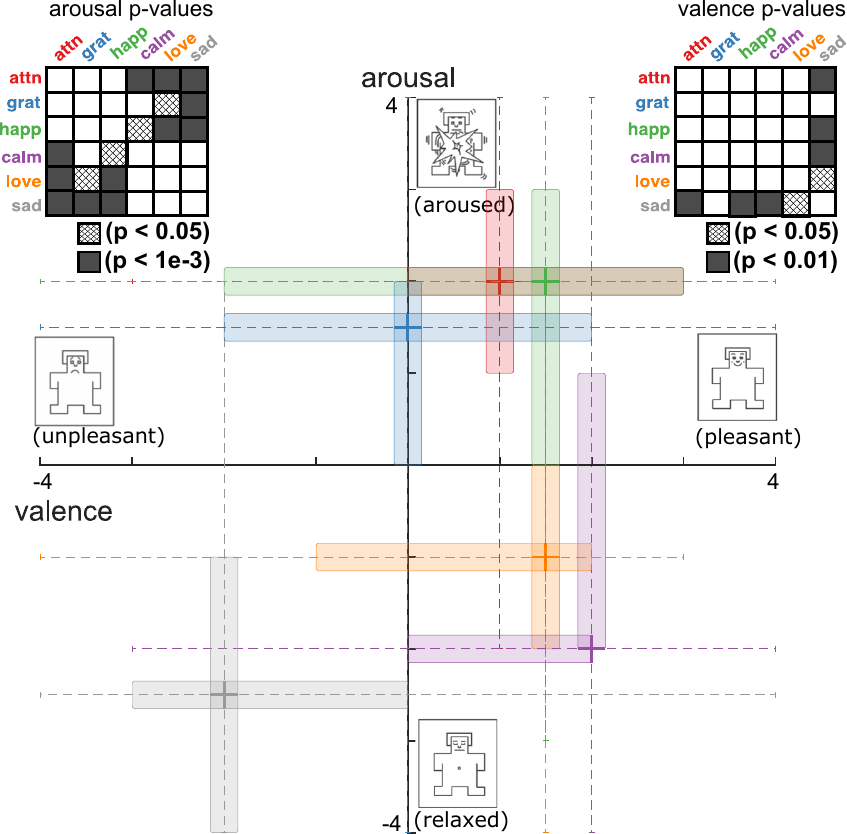}
  \vspace{-\baselineskip}
  \caption{The valence and arousal ratings for each scenario across participants as box-and-whisker plots. The + symbol for each scenario represents the point of median valence and arousal ratings. To determine differences in arousal, we performed a Kruskal-Wallis test comparing responses per scenario. Scenario was a significant factor ($\chi^2(1, N=180) = 23.26$, $p = 3.0\textrm{e-}4$). A similar test was performed for valance, and scenario was a significant factor ($\chi^2(1, N=180) = 64.16$, $p = 1.7\textrm{e-}12$). The two tables show post-hoc p-value levels for each pair of scenarios using the Dunn-Sidak test. We choose our tests as the data was not normal or equivarient, but samples were independent.}
  \label{fig:manikin}
  \vspace{-\baselineskip}
\end{figure}
\begin{figure*}[t!]
  \includegraphics[width=\textwidth]{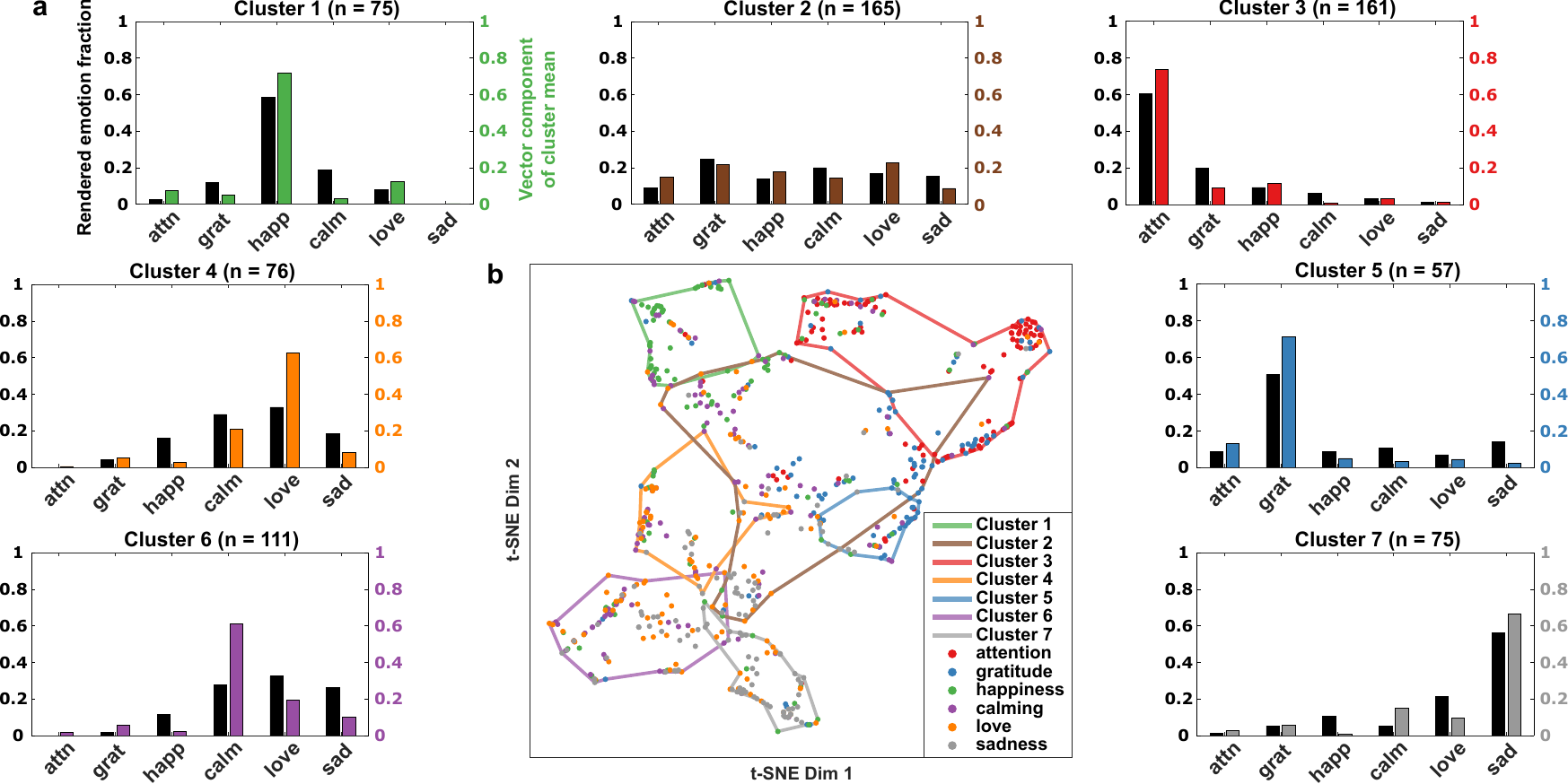}
  \vspace{-\baselineskip}
  \caption{Probability assignment cluster analysis. Each participant assigned the probability that they felt a given touch was drawn from each of the 6 scenarios displayed to them, providing a 6-dimensional vector with unit L1 norm. Outer figures (\textbf{a}) K-means analysis. The points were clustered to find consistent perspectives across participants. The black bars show the fraction of data points in the cluster where the signal played was from that scenario. The colored bars show the vector value of the mean of the cluster. This can be interpreted as the black bar showing the presented signal, and colored bar showing the likelihood participants in that cluster attributed the signal to that scenario. (\textbf{b}) t-SNE visualization. We performed t-SNE visualization on all data points to map them to 2D. The point color represents the presented signal. We then drew a hull of the points in each cluster from (a). This provides a two-dimensional visualization of how clusters border each other, in order to provide intuition for what the graphs in (a) represent.}
  \label{fig:kmeans}
  \vspace{-\baselineskip}
\end{figure*}

\subsection{Results}

Fig.~\ref{fig:top1}a shows a confusion matrix of the highest probability (``top choice'') scenario selected by participants for each trial in the scenario classification experiment. We found participants identified the scenario the signal was drawn from with overall accuracy of 45\% (2.7 times the rate of chance). For comparison, Fig.~\ref{fig:top1}b shows the results of a prior study on single-choice scenario classification based on direct human-human touch interaction~\cite{hauser2019uncovering}, which had overall accuracy of 57\% (3.4 times the rate of chance). For each row of each confusion matrix we normalize the values to 1 to calculate the Bhattacharrya coefficient~\cite{bhattacharyya1946} between the machine-human and human-human row values (Fig.~\ref{fig:top1}c). The attention, love, and sadness responses have larger coefficients than the gratitude, happiness, and calming responses. Additionally, Supplementary Information: Scenario Classification Experiment provides scenario classification top choice results confusion matrix for the first round of responses only for comparison.

Fig.~\ref{fig:manikin} shows the valence/arousal results in a circumplex affect model graph~\cite{Russell1980ACM}. The medians for sadness, calming, and happiness exist in the same quadrants as the foundational work in the area~\cite{Russell1980ACM} (love, attention, and gratitude do not appear in~\cite{Russell1980ACM}).

Each participant assigned a probability that the perceived touch was drawn from each of the six scenarios. Fig.~\ref{fig:kmeans}a shows a data clustering based analysis using the probability assignments. Each assignment can be represented as a 6-dimensional vector with unit L1 norm. We used k-means clustering to find groups of similar response vectors across participants. From both the Calinski-Harabasz~\cite{Caliski1974ADM} and Silhouette~\cite{Rousseeuw1987SilhouettesAG} clustering evaluation criteria, we found the optimal number of clusters to be seven. Clusters 1, 3, 5, and 7 indicate that a large number of participants assigned each of happiness, attention, gratitude, and sadness with high probability when presented with the respective signal. Cluster 4 shows that some participants assigned high probability for love when the true signal was calming or love, and to a lesser extent happiness or sadness. Cluster 6 shows that some participants assigned high probability for calming when the true signal was calming, love, or sadness. Cluster 2 represents high uncertainty. These clusters indicate that while many participants were certain of attention, gratitude, and happiness, participants also responded that each of calming, love, and sadness were likely to be calming or love in many cases. Fig.~\ref{fig:kmeans}b shows a 2D t-Distributed Stochastic Neighbor Embedding (t-SNE)~\cite{Maaten2008VisualizingDU} plot of the scenario vectors, with a hull of each cluster's points drawn. This plot provides another means to visualize data from Fig.~\ref{fig:kmeans}a, as well as intuition on which clusters border each other. Similar clustering analysis is provided on Self-Assessment Manikin data in Supplementary Fig.~S5.

The results show that skin stimulation using only eight one-degree-of-freedom actuators can convey social touch scenario information. Our cluster analysis shows where distinguishing information between signals is possibly lost. 

\section{Discussion}

\subsection{Similarity of Machine-Human Scenario Classification with Human-Human Interaction}

With no training, humans achieved high scenario classification accuracy using our system: 45\% for 6 classes, as compared to 57\% accuracy of human-human interaction shown in prior research~\cite{hauser2019uncovering} and 16.7\% for chance. In contrast to~\cite{hauser2019uncovering}, our experiment was performed using subjects who had no known relationship to the person from whom the signal was generated. Research has shown that individuals in romantic relationships are more accurate at distinguishing emotions than pairs of strangers~\cite{thompson2011effect}, so signals generated from a partner could potentially achieve higher accuracy, and a lower accuracy than~\cite{hauser2019uncovering} is not unexpected. Additionally, we rendered all signals on the forearm, including those sensed from the upper arm, which may have hindered the naturalness of some signals.

Our results show that it is possible to create signals using natural human expressions for detailed haptic communication. Using Fig.~\ref{fig:top1} we can compare patterns of recognition between our results and those of human-human touch~\cite{hauser2019uncovering}. We use the Bhattacharyya coefficient~\cite{bhattacharyya1946} as a measure to compare machine-human and human-human response similarity (Fig.~\ref{fig:top1}c). We observe that the response patterns for love, sadness, and attention have higher Bhattacharyya coefficient than the response patterns for gratitude, happiness, and calming. In particular, for true signal gratitude, our results involve less relative confusion between happiness and gratitude, and a higher rate of confusion between attention and gratitude. Our particular prompts likely influenced these confusion patterns. Our gratitude scenario (Supplementary Information: Social Touch Dataset), prompted subjects to touch surreptitiously. We believe this is why the measured gratitude signals are short (Fig.~\ref{fig:dataset}d), resulting in short press/squeeze actions which could be confused with attention. This highlights the importance of providing context to more fully understand social touch data and create signals for remote communication. 

Cluster analysis can determine common scenario probability assignments for each signal. If all participants drew from the same distribution of assignments, we would expect one cluster for each scenario, with weights equal to the corresponding row of the confusion matrix. However, we see cases where the cluster analysis contains information not in the top choice confusion matrix (Fig.~\ref{fig:top1}a). For example, in top choice analysis for true class gratitude, we observe a nearly equal number of examples for predicted class gratitude and attention. In the cluster analysis (Fig.~\ref{fig:kmeans}a), for true class gratitude, there is a cluster with high probability assignment for gratitude, but a relatively low probability assignment for attention. This implies that there was a class of people who were confident that gratitude, indeed, was gratitude. Based on the lack of a cluster with a large number of true class gratitude with high probability assignment for attention, we do not see a group of participants who were confident that gratitude was attention. This analysis demonstrates the importance of looking beyond top choice analysis, enabling detailed analysis of where subject confusions exist.

While our task was forced-choice, our probability assignment measure allows us to measure similar information to a non-forced-choice task. In the k-means clustering, Cluster 2 informs us of a group of participants that are likely to be uncertain of the best choice. In Supplementary Information: Scenario Classification Experiment we provide information on top-choice accuracy as a function assigned scenario probabilities to more explicitly simulate non-forced-choice.

\subsection{Implications of Sparse Representation}
The above results were achieved with a sparse display of haptic signals, using only eight, fixed-location, one-degree-of-freedom actuators placed at larger distances than the threshold for discrimination by afferents in the skin, based on direction discrimination tests~\cite{ackerley2014touch}. Our results support human the spatial sensitivity of afferents is higher than required to classify social touch signals.

Our work shows that short, sparse haptic signals can convey social touch information. Thus, such signals could be used to create a set of haptic icons, similar to text emojis, for sending social messages. We expect that these signals would be combined with other means of communication, such as video chat, video of a partner interacting via a virtual avatar, or integration with text communication. Given that touch can feel more or less pleasant depending on who administers the touch \cite{gazzola2012primary}, our signals may be able to convey even more salient affective significance through combination with vision and sound. In addition, individuals can use the algorithm to create haptic icons specific to their partners, for more personalized affective expression.

\subsection{Data-Driven Rendering}
Our work shows the benefit of a data-driven approach to social communication on simple array hardware. These contributions could lead to cheap, lightweight, wearable consumer social touch devices, or integration of such devices into more complex systems with minimal extra cost and weight. This is in contrast to current consumer haptic devices such as successors to the PHANTOM haptic device~\cite{massie1994phantom} (3D Systems, Rock Hill, SC, US) and the HaptX glove~\cite{haptx} (Seattle, WA, US), which are heavy, have limited workspaces, and are expensive. The algorithm is not specific to our dataset or hardware, so it can be applied to other types of actuators, including microfluidic haptic arrays \cite{besse2018large}, pin arrays \cite{Siu2018shapeShift2S}, and ultrasonic devices \cite{carter2013ultrahaptics}. Our dataset is also freely available to be applied for further research.

\subsection{Use of Scenarios as Social Touch Prompt}
To generate social touch data, we chose to provide specific scenarios rather than discrete affective category labels for the following two reasons: First, we wished to maximize the probability that the subjects performed authentic touches. Genuine emotion has been shown to be distinguishable from non-genuine emotion for facial expressions \cite{gosselin1995components}. Based on verbal feedback, informal pilot testing indicated that subjects were engaged by the scenarios. Second, we wished to minimize the impact of cultural perception, language ambiguity, or personal history in the terms we used. Different cultures express different levels of valence and arousal to the same emotional keywords \cite{bann2013measuring}. In general, these terms also have broad definitions, capable of encompassing a spectrum of affective states that could depend heavily on each user at a given time. Therefore, we posit that in our dataset the exact nature of what someone is trying to express is clearer than in datasets which assume that all participants have a similar interpretation of a single-word cue. For these reasons, we propose that the choice of scenarios over discrete affective category labels resulted in a less ambiguous task for those in the classification experiment. Further analysis on the impact of single-word cues compared to scenarios is an interesting direction for future work. 

We acknowledge that such scenarios are limited in their ability to capture a complex affective state. As mentioned, one use of this work could be to create haptic icons similar to emojis, which are limited in their precise expressive capability, but understandable in context.

\section*{Acknowledgments}

Research reported in this publication was supported by Facebook, Inc. The authors acknowledge additional support for S.R.W. and C.M.N. from the National Science Foundation Graduate Research Fellowship Program. We thank the behavioral scientists at IDEO who contributed to study design.

\bibliographystyle{IEEEtran}

\bibliography{CHARMBib, SocialBib}

%

\vspace{-12mm}

\begin{IEEEbiography}[{\includegraphics[width=1in,height=1.25in,clip,keepaspectratio]{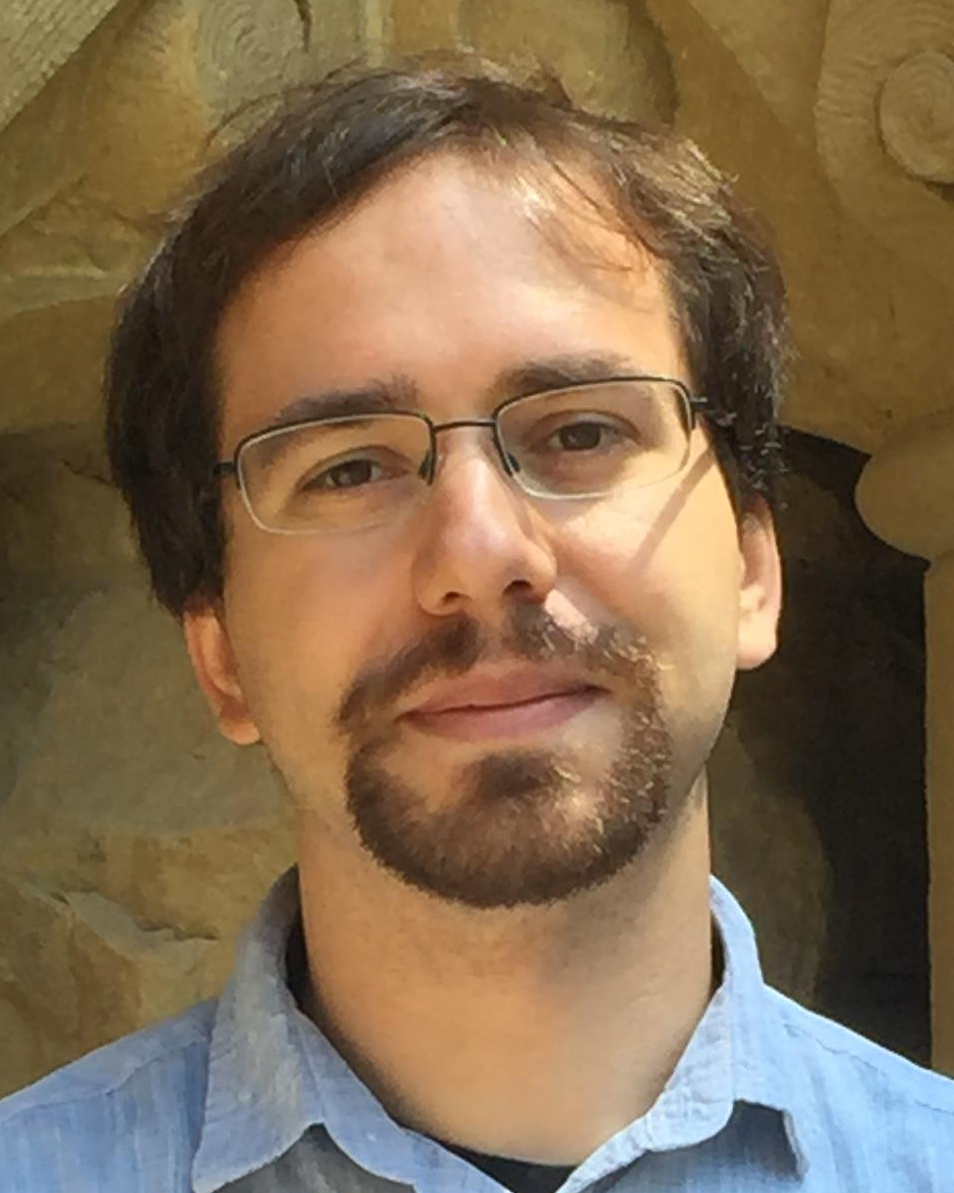}}]{Mike Salvato}
received their S.B.\ and M.Eng.\ in computer science from the Massachusetts Institute of Technology, where they conducted research with the Marine Robotics Group. They are currently a PhD candidate in mechanical engineering at Stanford University. Prior to their PhD they worked as a research engineer at the Allen Institute for Artificial Intelligence. Their research on haptics and computer vision aims to improve sensory feedback during interactions in virtual and augmented reality.
\end{IEEEbiography}

\vspace{-12mm}

\begin{IEEEbiography}[{\includegraphics[width=1in,height=1.25in,clip,keepaspectratio]{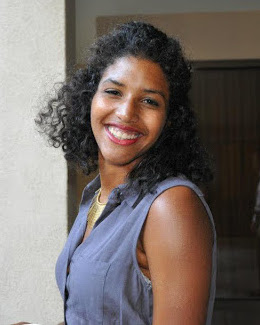}}]{Sophia R. Williams} received her B.S.\ from Harvey Mudd College in 2015 and her M.S.\ and Ph.D.\ from Stanford University in 2019 and 2021, respectively. Her research interests include wearable haptic interfaces, psychophysics, design, robotics, and human-robot interaction. Sophia is a recipient of the National Science Foundation Graduate Research Fellowship, Stanford Robotics Center fellowship sponsored by FANUC, and Watson Fellowship. She now works at Auris Health.
\end{IEEEbiography}

\vspace{-12mm}

\begin{IEEEbiography}[{\includegraphics[width=1in,height=1.25in,clip,keepaspectratio]{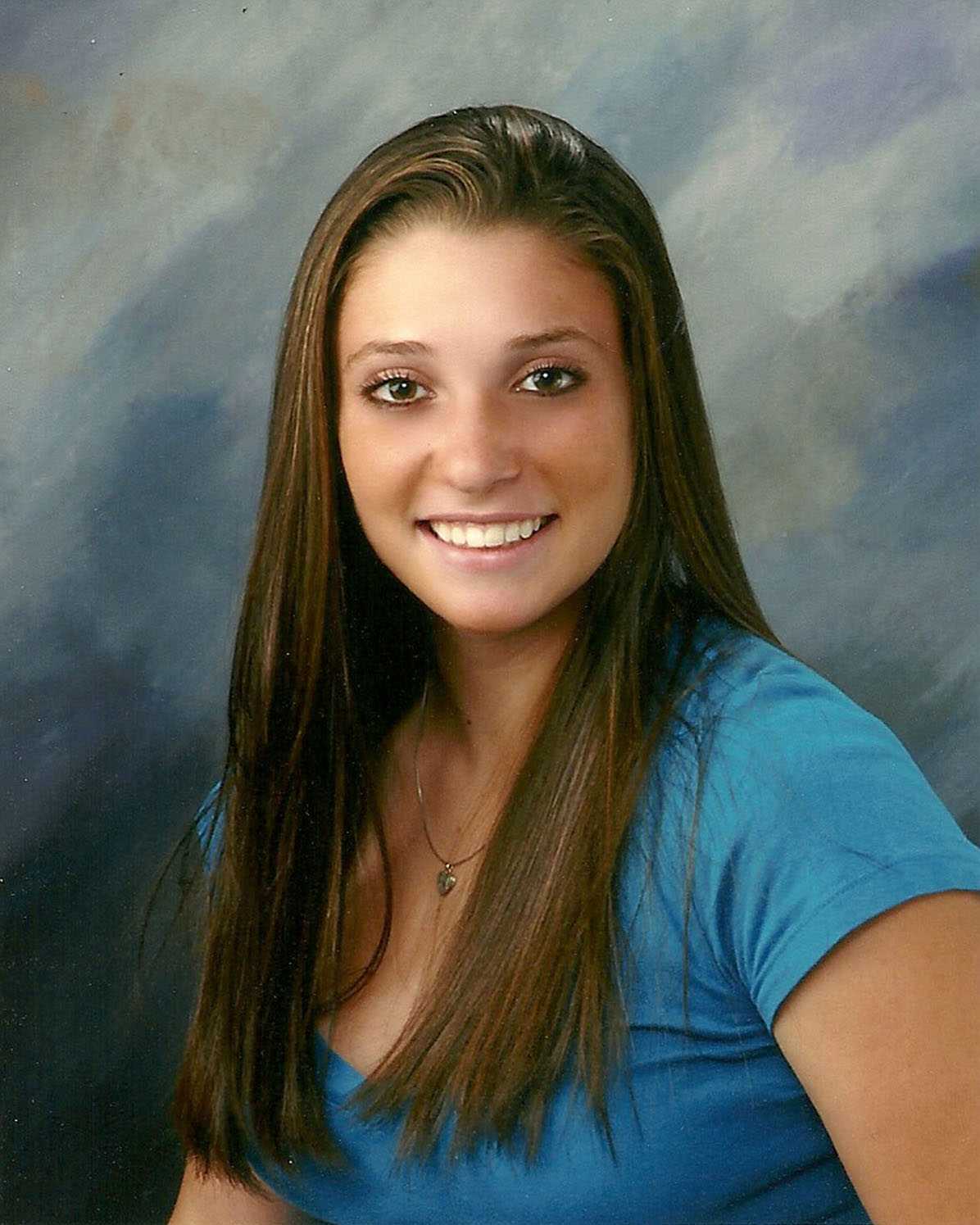}}]{Cara M. Nunez} received a B.S.\ in biomedical engineering and a B.A.\ in Spanish from the University of Rhode Island in 2016, the M.S.\ degree in mechanical engineering from Stanford University in 2018, and a Ph.D.\ in bioengineering at Stanford University in 2021. She was a National Science Foundation Graduate Research Fellow at Stanford and a DAAD Graduate Research Fellow at the Max Planck Institute for Intelligent Systems. Her research interests include haptic perception, cutaneous feedback, and wearable devices for medical applications, human-robot interaction, virtual reality, and STEM education.
\end{IEEEbiography}

\vspace{-12mm}

\begin{IEEEbiography}[{\includegraphics[width=1in,height=1.25in,clip,keepaspectratio]{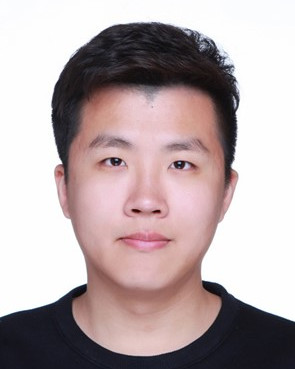}}]{Xin Zhu}
received his BSc in Computer Science from Beijing Jiaotong University, Beijing, China, and MS in Computer Science from University of Southern California in 2019. He is currently a PhD student in HaRVI lab, USC. Xin's research interests include haptics technology with social touch, virtual reality and HCI.
\end{IEEEbiography}

\vspace{-12mm}

\begin{IEEEbiography}[{\includegraphics[width=1in,height=1.25in,clip,keepaspectratio]{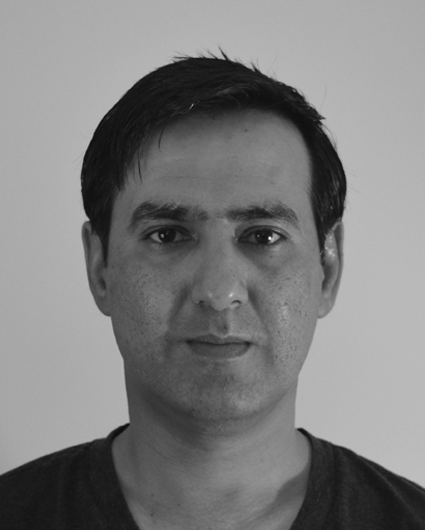}}]{Ali Israr}
received his B.Sc.\ in Mechanical Engineering from University of Engineering \& Technology, Lahore, Pakistan, and M.S/\ and Ph.D.\ in mechanical engineering from Purdue University in 2004 and 2007, respectively. He was at Rice University as postdoc and has since led haptic projects in Disney Research and Facebook, leading efforts to expand the utility of haptic enabled technologies for entertainment, educational, social, therapeutic and ARVR settings.
\end{IEEEbiography}

\vspace{-12mm}

\begin{IEEEbiography}[{\includegraphics[width=1in,height=1.25in,clip,keepaspectratio]{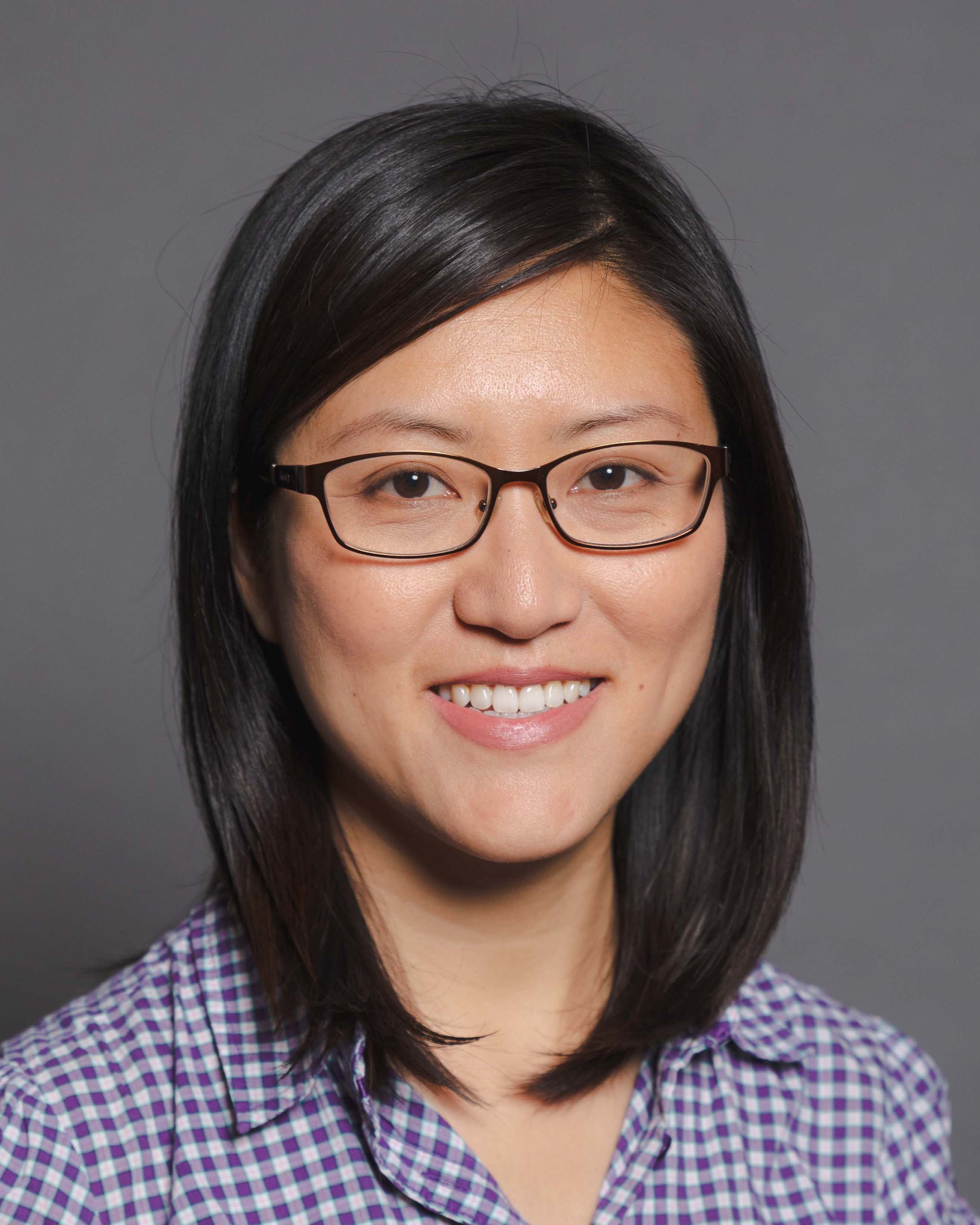}}]{Frances W. Y. Lau}
received the B.A.Sc. degree in electrical engineering from the University of Toronto in 2005, and the M.S.\ degree and Ph.D. degrees in electrical engineering from Stanford University in 2007 and 2013, respectively.  Since 2016, she has been working at Facebook on research and development of systems that act as interfaces between humans and technology, using methods that range from haptics to brain-computer interfaces. 
\end{IEEEbiography}

\vspace{-12mm}

\begin{IEEEbiography}[{\includegraphics[width=1in,height=1.25in,clip,keepaspectratio]{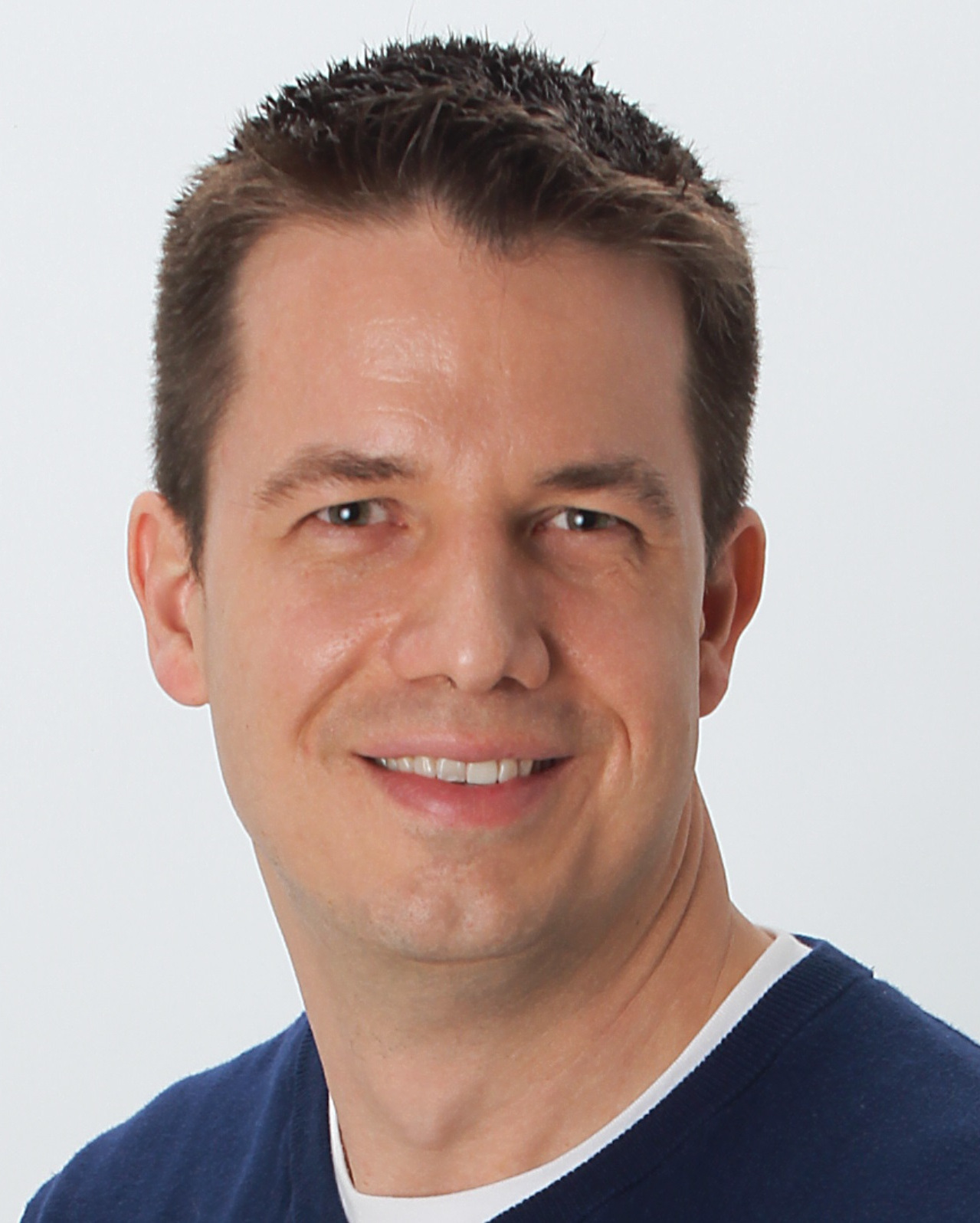}}]{Keith Klumb}
completed his BS at Purdue University and is currently a Technical/Research Program Manager. He has worked on research and new technology initiatives including BCI, Soli Radar, haptics psychophysics and HR monitoring at various companies including Google, Facebook and Fitbit. 
\end{IEEEbiography}

\vspace{-12mm}

\begin{IEEEbiography}[{\includegraphics[width=1in,height=1.25in,clip,keepaspectratio]{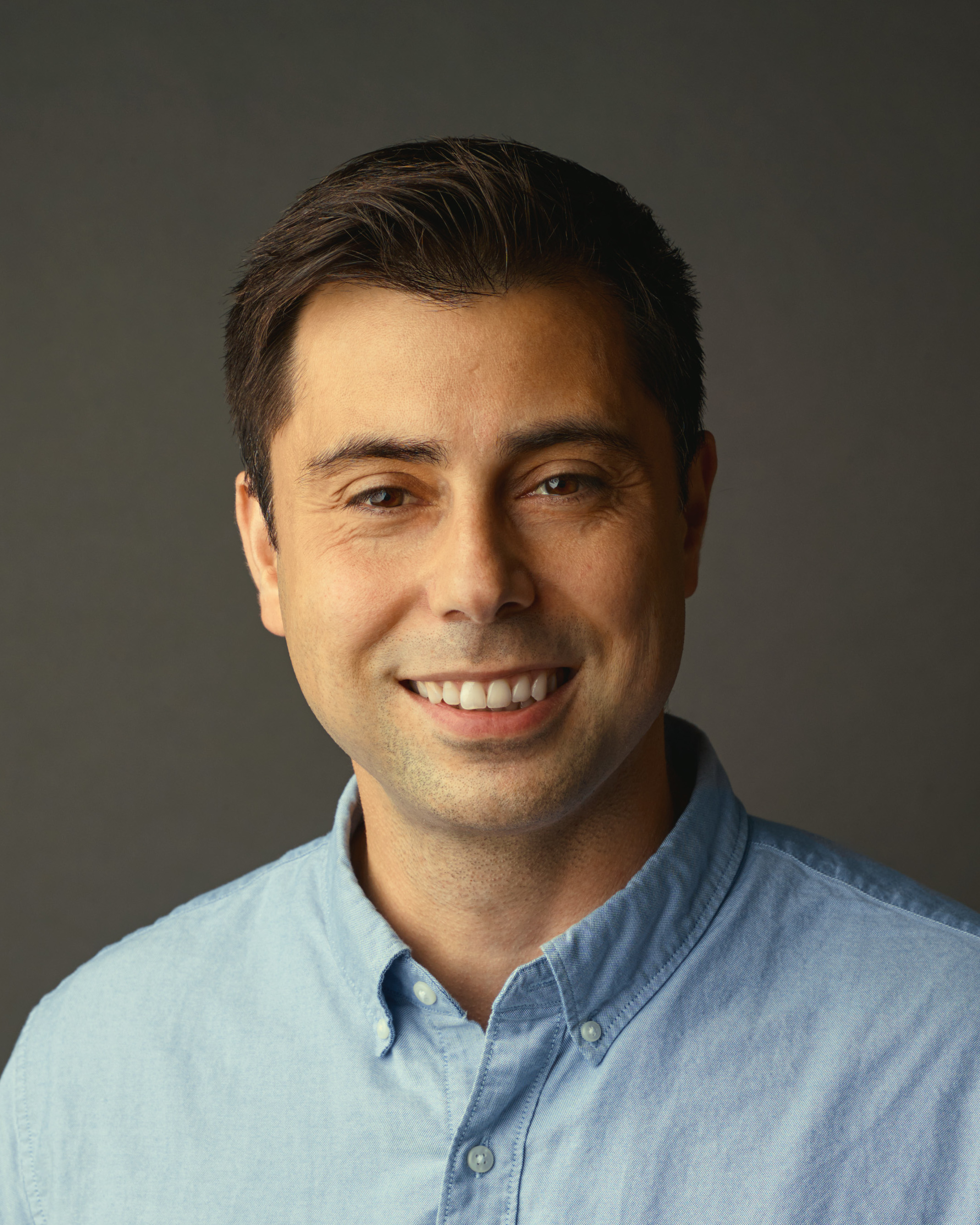}}]{Freddy Abnousi}
is the Head of Health Technology at Facebook and a practicing Interventional Cardiologist. He also serves as Innovation Advisor to the American College of Cardiology, Professor Adjunct at Stanford University School of Medicine, and Assistant Professor Adjunct at Yale University School of Medicine. 
He completed Fellowships in Cardiovascular Medicine and Interventional Cardiology, as well as Residency in Internal Medicine at Stanford University Medical Center. He was previously a resident surgeon at the University of California, San Francisco. He completed his MD at Stanford University School of Medicine, MBA from Oxford University, and MSc in Health Policy, Planning, \& Financing from the London School of Economics. 
\end{IEEEbiography}

\vspace{-12mm}

\begin{IEEEbiography}[{\includegraphics[width=1in,height=1.25in,clip,keepaspectratio]{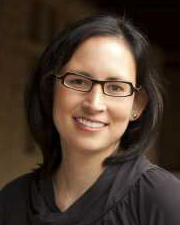}}]{Allison M. Okamura}
received the B.S. degree from the University of California, Berkeley, in 1994, and the M.S. and Ph.D. degrees from Stanford University, Stanford, CA, in 1996 and 2000, respectively, all in mechanical engineering. She is currently a Professor of mechanical engineering at Stanford University, Stanford, CA. Her research interests include haptics, teleoperation, medical robotics, virtual environments and simulation, neuromechanics and rehabilitation, prosthetics, and engineering education.
\end{IEEEbiography}

\vspace{-12mm}

\begin{IEEEbiography}[{\includegraphics[width=1in,height=1.25in,clip,keepaspectratio]{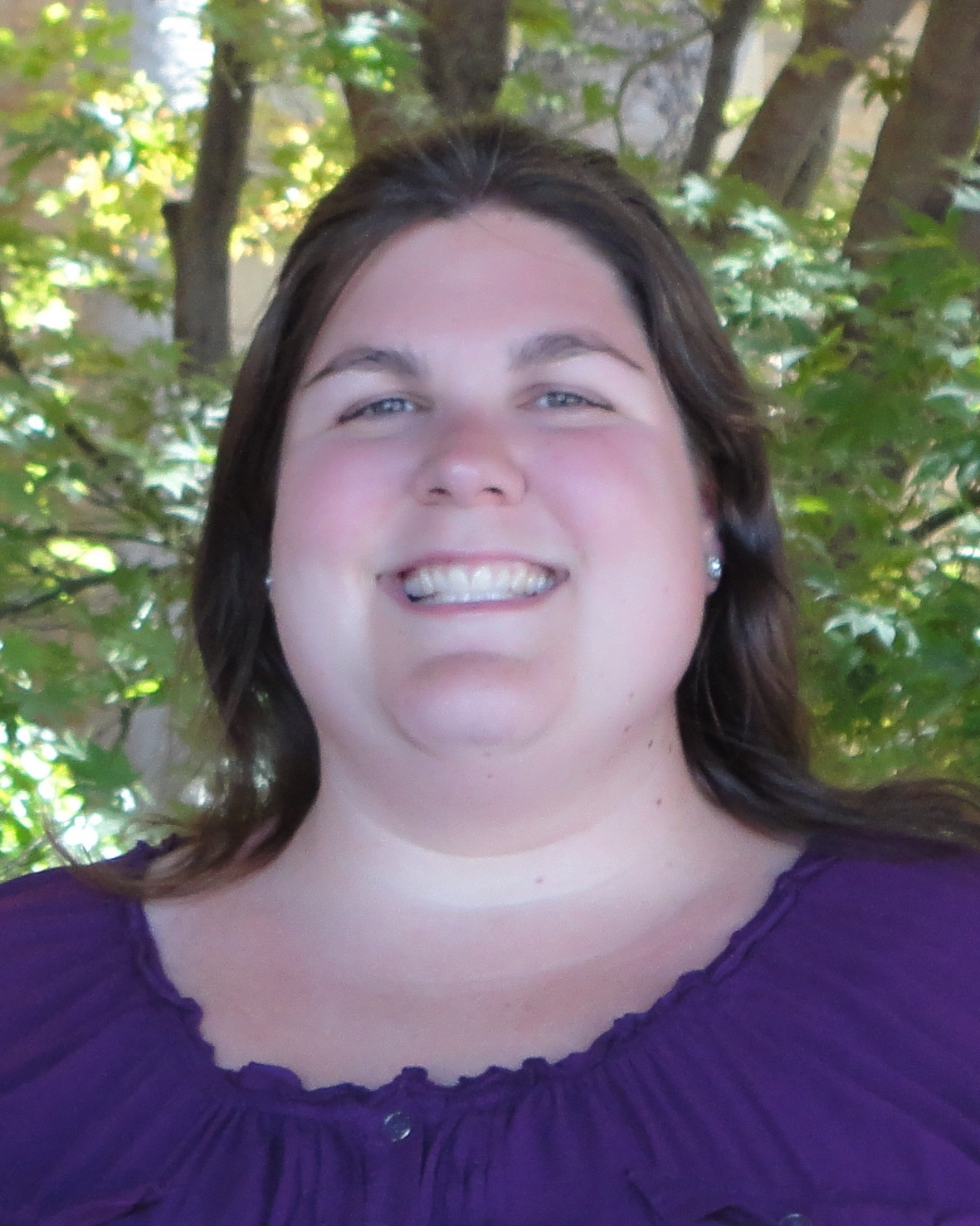}}]{Heather Culbertson}
Heather Culbertson is an assistant professor of Computer Science and Aerospace and Mechanical Engineering at the University of Southern California. She received the M.S.\ and Ph.D.\ degrees in the department of mechanical engineering and applied mechanics (MEAM) at the University of Pennsylvania in 2013 and 2015, respectively. She earned the B.S. degree in mechanical engineering at the University of Nevada, Reno in 2010. Prior to joining USC she was a research scientist at Stanford University. Her research focuses on the design and control of haptic devices and rendering systems, human-robot interaction, and virtual reality.
\end{IEEEbiography}







\end{document}


%
\title{Data-driven sparse skin stimulation can convey social touch information to humans\\ \rule{\textwidth}{.6pt} \vspace{-3mm} \\Supplementary Information}
%
%
%
%

\author{Mike Salvato, Sophia R. Williams, Cara M. Nunez, Xin Zhu, Ali Israr, Frances Lau, Keith Klumb,\\ Freddy Abnousi, Allison M. Okamura, Heather Culbertson

}
\maketitle

\IEEEdisplaynontitleabstractindextext

%
\IEEEpeerreviewmaketitle

\IEEEraisesectionheading{\section{Social Touch Dataset}\label{si:socialtouch}}

\subsection{Scenario Prompts}
The scenario prompts that the toucher and receiver are given are shown below. The amusement prompt was used to acclimate participants with the procedure. The toucher audios were used in the scenario classification experiment. See Dataset~SD1 for recordings of these prompts.

\subsubsection*{Amusement (Used for training only)}
\textit{\indent Toucher:} Imagine this: You're spending time with the person sitting next to you. You're talking about small things here and there, but there are also some long pauses where neither of you has much to say. Then, they make a face and say something really, really funny. Suddenly everything feels lighter. You reach out and touch them to show your delighted amusement.

\textit{Receiver:} Imagine this: You're spending time with the person sitting next to you. You're talking about small things here and there, but there are also some long pauses where neither of you has much to say. Sure, it’s a little awkward. Silence is fine, but today you’re feeling chatty. The person sitting next to you seems receptive, so you crack a joke. It must have worked, because suddenly everything feels lighter. When they touch you, expressing their amusement, you feel like maybe you saved the day, or at least the next ten minutes.

\subsubsection*{Attention}
\textit{\indent Toucher:} Imagine this: You're at a crowded party with the person sitting next to you, and they've drifted into a conversation with someone else. You don't want to be rude and interrupt them, but you really need to ask them an important question. Turn back toward them, and touch them in a way that gets their attention.

\textit{Receiver:} Imagine this: You're at a crowded party with the person sitting next to you, but you've drifted off into a side conversation with other people. The person you're talking to is telling you a fascinating story, and you're completely rapt. It's like no one else is in the room. You're not purposefully ignoring the person you came to the party with, but you're really focused on hearing all the details. Then, the person you came to the party with touches you in a way that signals they need your attention.

\subsubsection*{Gratitude}
\textit{\indent Toucher:} Imagine this: You and the person sitting next to you are at a dinner party with a group of friends you've known for a long time. You start to sense the conversation getting dangerously close to a topic that could put you in a really uncomfortable position if aired in this group. The person sitting next to you picks up on your uneasiness and deftly takes the conversation in a different direction. You feel a deep sense of relief, and you want them to know you are grateful. Reach out and touch this person to express your gratitude.

\textit{Receiver:} Imagine this: You and the person sitting next to you are at a dinner party with a group of friends you've known for a long time. You have a knack for being able to spot a train wreck before it happens, so when you sense the conversation is getting dangerously close to a topic that could damage the reputation of the person sitting next to you, you steer it back on track. You don't expect any credit for such agile social maneuvers, but when they reach out and telegraph ``thank you'' with their touch, you instantly know you've done good.

\subsubsection*{Happiness}
\textit{\indent Toucher:} Imagine this: Today is the perfect day. Like magic, everything is going right, and everyone around you seems to be in a good mood. You're walking down the street in the sunshine, listening to your favorite song. Feel that little bounce in your step? You catch a glimpse of yourself in a store window, and dang -- you look good! On a whim, you decide to pop in and buy a lottery ticket. Why not? Scratch, scratch, scratch...you win \$50! This day just couldn't get any better. Now take this feeling of happiness and touch the person next to you to express it.

\textit{Receiver:} Imagine this: When the person walking toward you just can't stop smiling, you know something is going really right. They're beaming, and you can just tell they're having the best day ever, almost walking on clouds. They bound over to you and reach out to touch you, and it's like an electric bolt of pure joy flows through you.

\subsubsection*{Calming}
\textit{\indent Toucher:} Imagine this: It's 7:30 pm on a miserable, rainy Thursday, and you're waiting for the person next to you to meet you for dinner. It's been one hell of a week for them. Everything that could go wrong for them has gone wrong. Finally, you see them walk through the door, and they’re completely frazzled. You can practically feel the stressed out energy radiating off their body. Go to this person and touch them in a way that feels calming.

\textit{Receiver:} Imagine this: What a crappy week. You're stressed out, and things just keep piling on. You really aren't in the mood to meet the person sitting next to you for dinner, but you can't back out now. You walk through the door in a state, and that must show because they reach out and touch you in the most compassionate and tender way. You feel instantly understood, and it brings a wave of calm. Your blood pressure feels like it just dropped 20 points, in a good way.

\subsubsection*{Love}
\textit{\indent Toucher:} Imagine this: You and the person sitting next to you are spending the afternoon together. You're walking to get a bite, the weather is amazing, and you're catching up on everything in the way that friends do. You look at them, and it suddenly strikes you how much this friendship means to you, that life is so much easier and better with them around. Reach out and touch this person to express your love for them.

\begin{figure*}[t!]
\centering
  \includegraphics[width=\textwidth]{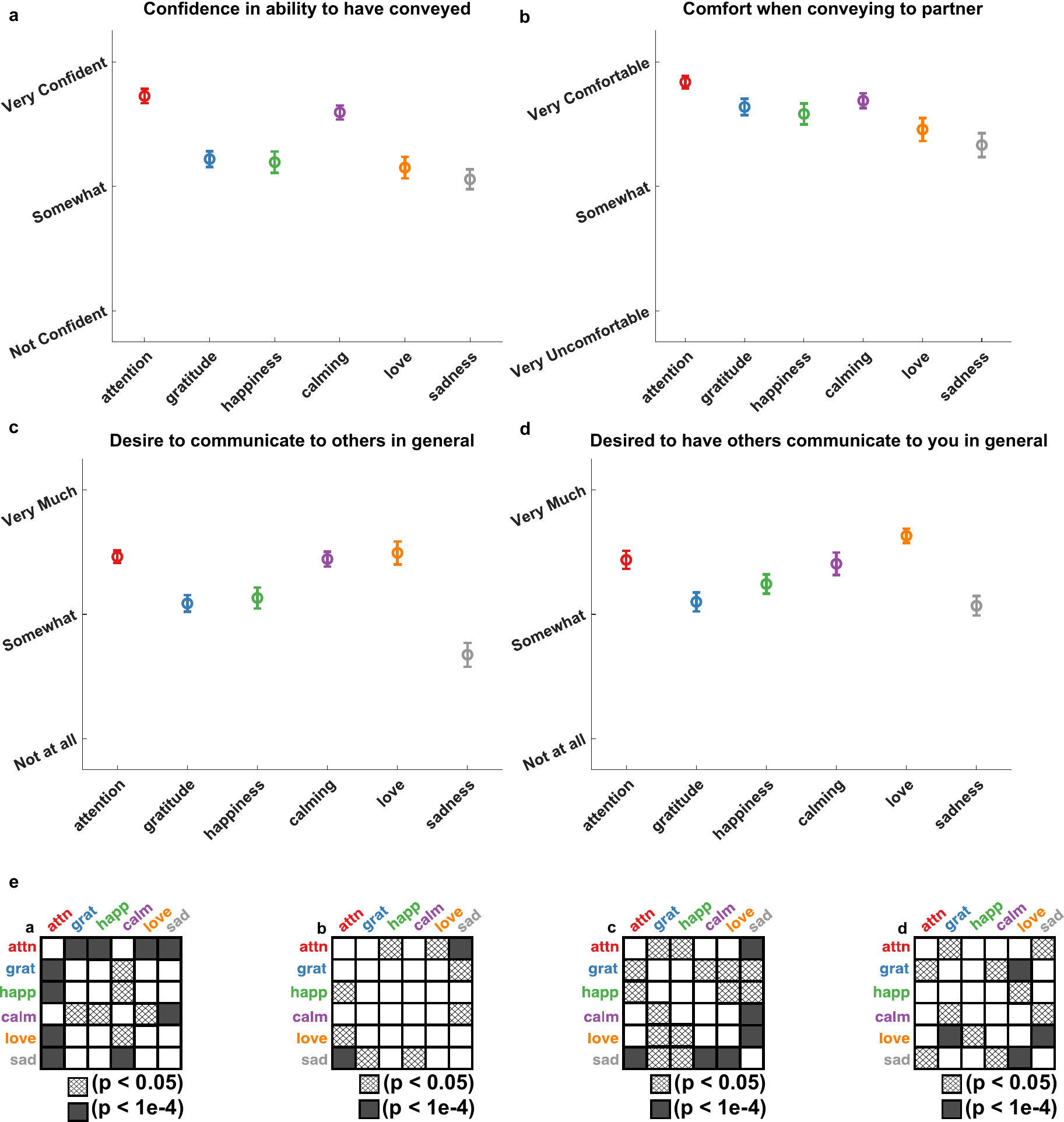}
  \caption{Social touch recording survey results. (\textbf{a}-\textbf{d}) The survey results for each question, based on a 7-point Likert scale. Error bars indicate standard error. (\textbf{e}) Associated significance matrix for each question, using the Dunn-Sidak post-hoc test for Friedman. This test was chosen as our data was non-normal and not independent.}
  \label{sif:survey}
\end{figure*}

\textit{Receiver:} Imagine this: You and the person sitting next to you are spending the afternoon together. You're walking to get a bite, the weather is amazing, and you're catching up on everything in the way that friends do. You look at them, and it suddenly strikes you how much this friendship means to you, that life is so much easier and better with them around. They reach out to express their love for you.

\subsubsection*{Sadness}
\textit{\indent Toucher:} Take a moment to think about someone you have lost -- could be the death of someone close, or a breakup that tore you apart. Sit with that feeling for a bit... locate it in your body. Maybe it feels heavy, or achy. Get in touch with the sadness you feel about this loss, and touch the person next to you in a way that expresses that sadness.

\textit{Receiver:} Even if it doesn't say anything out loud, a heavy heart is a loud presence. It's almost like another person in the room. Or maybe instead of a presence what you are feeling is an absence. The absence of joy. It seems like the person next to you is in mourning, like they've lost something that was important to them. That makes you instantly want to fill up the space with something like compassion or help or just being there.

\subsection{Social Touch Survey}

We provided a survey in which we asked participants to rate their thoughts on conveying each meaning through touch during the data collection process and in general along four scales. We used the Friedman test to check the effect of touch meaning on response, as our data was non-normal and not independent. We provide the exact prompt for the "Attention" version of each question:

\begin{itemize}
    \item \textit{Rate your confidence in your ability to have conveyed \textbf{Attention} through touch alone.} (Fig.~\ref*{sif:survey}a) The Friedman test shows that touch meaning has a significant effect on participants' confidence in their ability to convey that meaning ($\chi^2(5) =$ 58.74, $p =$ 2.2\textrm{e-}11). The mean values for attention and gratitude were higher that the other touch meanings.
    
    \item \textit{Rate your comfort level when conveying \textbf{Attention} to your partner in this study.} (Fig.~\ref*{sif:survey}b) The Friedman test also shows that touch meaning has a significant effect on participants' comfort when conveying an meaning to a partner ($\chi^2(5) =$ 38.55, $p =$ 2.9\textrm{e-}4). The highest average comfort level was reported for attention, whose distribution was statistically significantly different from the distributions for love and sadness. 
    
    \item \textit{In general, how much do you want to communicate \textbf{Attention} to other people through touch?} (Fig.~\ref*{sif:survey}c) The Friedman test shows that touch meaning has a significant effect on participants' desire to communicate that meaning to others ($\chi^2(5) =$71.48, $p =$ 5.0\textrm{e-}14). Participants have the highest desire to communicate attention, calming, happiness, and love.
    
    \item \textit{In general, how much do you want others to communicate \textbf{Attention} to you through touch?} (Fig.~\ref*{sif:survey}d) The Friedman test shows that the touch meaning has a significant effect on participants' desire to have others communicate that meaning to them ($\chi^2(5) =$ 59.74, $p =$ 1.4\textrm{e-}11). The mean values indicate that participants most desire others to communicate love, attention, and calming to them.

\end{itemize}

\newpage~\newpage~\newpage

\newcommand{\CommentX}[1]{\unskip~~#1~}
\hypertarget{si:mapping}{\section{Mapping Algorithm Formalization}}

Here we provide a full formalization of our algorithm, which that maps from data recorded on a 2D sensor array to an array of actuators. Our use case is to map from a pressure sensor array recording of human touch to a device consisting of a series of actuators worn by a human. The algorithm tracks trajectories of high pressure datapoints whose pressures vary smoothly over space and time, and chooses which of those trajectories to map onto the actuators. For the trajectories selected, our algorithm chooses an accurate rendering given the the actuator workspace limitations. The code is provided in Code~SC1.

\subsection{Function Input and Output}

The algorithm provides a mapping \[g : f_1 ... \times ... f_p \times a_1 \times ... \times a_q \to \mathbb{R}^{3 \times p \times q}\] where $p$ is the number of data frames, $q$ is the number of actuators. $f_1, ..., f_p \in F$ are a series of data frames (which are not constrained to be rectangular) and  $a_1 ... a_q \in A$ are 3D geometric shapes.

Each frame $f_t \in F$ contains a series of coordinates $\mathbf{x} 
\in \mathbb{R}^2$ and associated intensities $I_t(\mathbf{x}) \in \mathbb{R}$ at time index $t$. These coordinates must be in metric coordinates (e.g. millimeters), and these frames define the ``coordinate space.'' The intensities provide a measure of the probability of detection at that coordinate.

Each shape $a_i \in A$ represents the workspace for actuator $i$. The workspace is defined by the actuator's available motions (e.g. up and down, side to side, etc.) and the extents of its motion in any direction. It is represented in the same units as the coordinate space. The algorithm tracks only the center of each actuator, so the shape should represent the bound of how far the actuator center can move.

For each frame for each actuator, a 3D value for the location of each actuator is output. Let $L_i(t) \in \mathbb{R}^{3}$ be the location of actuator $i$ at time $t$ in coordinate space. The first 2 dimensions of $L_i( t)$ are coordinate distances, and last dimension is the measurement value.

To map values from the algorithm to hardware, a function 
\[h : \mathbb{R}^{3} \to \mathbb{R}^{3}\] maps coordinate space actuator location values $L_i(t)$ to desired 3D locations for each actuator. Typically, the first two dimensions would be mapped by the identity function, representing no scaling between coordinate space and the actuator output in those dimensions. Because the intensity values of sensor frames $F$ are not necessarily directly mappable to a location in space, the third dimension will be determined based on the use case.

\subsection*{Step 1: Trajectory generation using multi-object tracking}

In the first phase of the algorithm, we leverage multi-object tracking algorithms~\cite{SupZhang2008GlobalDA} to find the optimal contiguous paths tracked in the sensor. Such algorithms are typically used in computer vision to extend object detectors from single images to continuously tracked objects in video. They rely on the object detector providing a probability of a correct detection to the tracking algorithm. They then take all detections across all images, and provide a consistent set of tracked object trajectories through the video. We use this method to locate areas of contiguous, high pressure readings of sensor, which we assume represent interaction of a participants hand on the sensor. Our method for trajectory generation is an adaption of~\cite{SupZhang2008GlobalDA} for our sensor frames. 

If we consider our sensor frames as video, we assume the pressure at each pixel location is monotonic with the probability that it represents the center of mass of pressure being applied by some object. This is a statement that higher pressure indicates a higher likelihood of meaningful contact with the sensor. Thus we could consider each pixel as a detection, with probability as a function of pressure. Let $\mathbf{g}$ be a pixel, where $\mathcal{G} = \{\mathbf{g}_1, ... \mathbf{g}_n\} = \{(t_1, \mathbf{x}_1), ..., (t_n, \mathbf{x}_n)\}$ is the set of all pixels, consisting of coordinates $\mathbf{x}$ over all times $t$, and let $V(\mathbf{g})$ be the intensity of a pixel. $V(\mathbf{g}) = V(t,\mathbf{x}) = I_t(\mathbf{x})$.  Let $\bar{V}$ be the mean intensity over all pixel values in the data and $\sigma_V$ be the standard deviation of those values. We use the following equations to calculate the functional probability of a detection:
\[z(\mathbf{g}_i) = \frac{V(\mathbf{g_i}) - \bar{V}}{\sigma_V}\]
\[P(\mathbf{g}_i) = \Phi(z(\mathbf{g}_i) - \sigma_k)*m\]
where $\Phi()$ is the cumulative distribution function (CDF) of the standard normal distribution. The above equation uses the z-score of a pixel value compared to all pixels in the entire sequence and determines its probability value by taking the standard normal CDF. $\sigma_k$ is used to determine the number of standard deviations from the mean that a pixel must be for a value to be considered to have 0.5 likelihood of being a detection -- in our case we used $\sigma_k=1.25$. The constant $m < 1$ is used to prevent undefined values in later steps. In our work we used $m=0.98$.

For computational efficiency we consider only pixels which are a local maxima. We define $\mathcal{G'} \subseteq \mathcal{G}$:
\[\mathcal{G'} =  \{\mathbf{g}_i \mid V(\mathbf{g}_i) \geq V(\mathbf{g}_j) \textrm{ for } | ||\mathbf{x}_i - \mathbf{x}_j || \leq \sqrt{2}\}\]
In addition, we need a measure of the transition probability between detections $P_{\textrm{link}}(\mathbf{g}_i | \mathbf{g}_j)$, where $t_i = t_j+1$
\[P_{\textrm{link}}(\mathbf{g}_i | \mathbf{g}_j) = \begin{cases} 1 - \frac{||\mathbf{x}_i - \mathbf{x}_j||}{k_d} & \text{if } ||\mathbf{x}_i - \mathbf{x}_j||\leq k_d \\ 0  & \textrm{otherwise} \end{cases}\]
where $k_d$ is a problem specific parameter. For our use case $k_d = 50$.

The goal is to find the set of trajectories $\mathcal{J}$ that best explains $\mathcal{G'}$~\cite{SupZhang2008GlobalDA}, with trajectories $J_k \in \mathcal{J}$, $J_k = \{\mathbf{g}_{k0}, ..., \mathbf{g}_{kl}\} \subset \mathcal{G'}$. This is formulated as a maximum a posteriori probability (MAP) estimate:
\[\mathcal{J}^* = \underset{\mathcal{J}}{\textrm{argmax }} P(\mathcal{J} \mid \mathcal{G'})\]
\[ = \underset{\mathcal{J}}{\textrm{argmax }} \prod_{i}{P(\mathbf{g}_i \mid \mathcal{J})} \prod_{J_k \in \mathcal{J}}P(J_k)\]
\[J_k \cap J_w = \emptyset, k \neq w\]
\[P(\mathbf{g}_i | \mathcal{J}) = \begin{cases} P(\mathbf{g}_i) & \mathbf{g}_i \in J_k  \in \mathcal{J} \\ 1- P(\mathbf{g}_i)  & \textrm{otherwise} \end{cases}\]
\[P(J_k) = P_\textrm{entr}(\mathbf{g}_{k0}) P_\textrm{link}(\mathbf{g}_{k1} | \mathbf{g}_{k0}) ... P_\textrm{link}(\mathbf{g}_{kl} | \mathbf{g}_{kl-1})P_\textrm{exit}(\mathbf{g}_{kl})\]

$P_\textrm{entr}$ is the probability that a point starts a trajectory, $P_\textrm{exit}$ is the probability a point ends a trajectory. $P_\textrm{entr} = P_\textrm{exit}$ is a tunable constant. We use $e^{-8}$, indicative of source/sink costs of 8 in the formulation by~\cite{SupZhang2008GlobalDA}. Increasing this value increases the number of trajectories that the algorithm is likely to find. 

The constraint $J_k \cap J_w = \emptyset, k \neq w$ indicates that two trajectories cannot share an element.

A min-cost flow framework to solve for $\mathcal{J}^*$ is presented in \cite{SupZhang2008GlobalDA}. 

\subsection*{Step 2: Actuator workspace restriction}

The previous section gave us $\mathcal{J}^*$ which represents a set of trajectories from the sensor data. We will now find a mapping from the trajectories to our actuators. 

Assume we have a set of 3D geometric shapes $A$ which represent the bounds of motion for the center of an actuator in trajectory space. Each vertex should be mapped into the coordinate space of our sensor frames $F$. By positioning the shapes onto a particular location on the frame, they represent some area of the sensor space that we wish to actuate. By trying many transforms of the set of shapes, we can find the optimal portion of the sensor space to render based on which trajectories pass through that area. Let $M$ represent the set of transforms we wish to try (e.g. translations and scalings). Let $A_m$ represent the set of actuator workspaces transformed by transform $m \in M$.

We then find the set of trajectories to render given a workspace restriction. We assume that a given actuator can only render one trajectory at a time, so we wish to find a set of trajectories that would not require any actuator to render more than one trajectory at any time $\tau \in T$, where $T = 1,...,p$ all times in which we have data frames. This implies a matching between two sets of data -- the set of trajectories at time $\tau$, and the set of actuator workspaces with any trajectories in them at time $\tau$. In order for a set of trajectories to be valid and maximal, there must then be a one-to-one correspondence between the two sets -- a bipartite perfect matching. Let 
\[\mathcal{J}^{*\tau} = \{J \in \mathcal{J}^* | \exists \mathbf{g}_i \in J, t_i = \tau\}\] Consider any $\Theta^{\tau} \subseteq \mathcal{J}^{*\tau}$. Let 
\begin{eqnarray*}\Xi^{\tau} = \{a \in m(A) | \exists J \in \Theta^{\tau} \textrm{ s.t. } \exists \mathbf{g}_i \in J, t_i = \tau, \\ \mathbf{x}_i \textrm{ within } a \in m(A)\}\end{eqnarray*} i.e. the actuator workspaces $a$ where there exists a pixel $ \mathbf{g}$ in a trajectory $J \in \Theta^\tau$, with location $\mathbf{x} \textrm{ within } a$ at time $\tau$. Consider the bipartite graph with vertex sets $\Theta^{\tau}$ (trajectories) and $\Xi^{\tau}$ (actuator workspace) and edges between them. In order for all $J\textrm{ s.t. }J \in \Theta^{\tau}$ to be compatible, the edges must be a bipartite perfect matching in this graph, otherwise two trajectories are in the same actuator workspace at the same time. By checking if such a bipartite perfect matching exists for each $\Theta^{\tau}$ and associated $\Xi^{\tau}$ for all $\tau \in T$, we can determine which subsets of $\mathcal{J}^*$ can be rendered.  For every timestep $\tau$, we consider the bipartite graphs for all possible all $\Theta^{\tau} \subseteq {J}^{*\tau}$ and associated $\Xi^{\tau}$. If a bipartite perfect matching is not possible for $\Theta^{\tau}$ we know the set of trajectories $J\textrm{ s.t. }J \in \Theta^{\tau}$ is invalid. Thus we can calculate the set of all sets of trajectories which are invalid $\Theta^\vee$. We use Hall's theorem to check the bipartite perfect matchings~\cite{SupChawla09}.

We wish to find the best set of compatible trajectories for our frame sequence and actuator workspace constraints. To do this we create a convex measure of trajectory quality: 
\[R(J, m(A)) = \sum_{\mathbf{g}_i \in J}{\log\left(\frac{P(\mathbf{g}_i)}{1-P(\mathbf{g}_i)}*D(\mathbf{g}_i, m(A)\right)}\] 
$D()$ is a small factor that weights trajectory elements lower as they move further from the center of the actuator bounding shape they are within at a given time. In this work we set 
\[D(\mathbf{g}_i, m(A)) = 1.02\left(1.04 - \frac{0.04||\mathbf{g}_i-c_i||_2^2}{c_r^2}\right)\] where $c_i$ is the center of the actuator workspace that $\mathbf{g}_i$ is inside of, and $c_r$ is the radius of the workspace. This factor assumes that it is more preferable to render the center of a workspace than the edges. Let $\Theta^\vee$ be the set of all sets of trajectories that are incompatible as above, and $\mathcal{J}^{*'}$ be the power set of $\mathcal{J}^*$. Let $k(J, \mathcal{J}) = 1$ iff $J \in \mathcal{J}$.  

We can then find the optimal trajectories by the following convex optimization problem:
\[\underset{m \in M}{\textrm{argmax }} \underset{\mathcal{J} \in \mathcal{J}^{*'}}{\textrm{argmax }} \underset{J \in \mathcal{J}}{\sum} R(J, m(A))\]
\[\textrm{subject to: } \left[\underset{J \in \mathcal{J}, J \in \Theta}{\sum} k(J, \mathcal{J})\right] < |\Theta|, \forall \Theta \in \Theta^\vee\]

The above can be solved with an off-the-shelf convex optimization solver such as CVX \cite{Supcvx,Supcvx2} for the inner optimization, and by iterating over all $m \in M$ for the outer optimization.

We could have chosen to merge Step 1 and Step 2 into a single optimization where the trajectories are selected with information about the actuator workspaces provided. However, we chose to separate these procedures so that Step 1 represents the trajectories strictly as a function of the data, so they are a more accurate representation of the data itself than if a single optimization were performed.

\clearpage
\newpage

\hypertarget{si:classification}{\section{Scenario Classification Experiment}}

\subsection{Non-forced-choice Simulation}

\begin{figure}
\centering
  \includegraphics[width=\columnwidth]{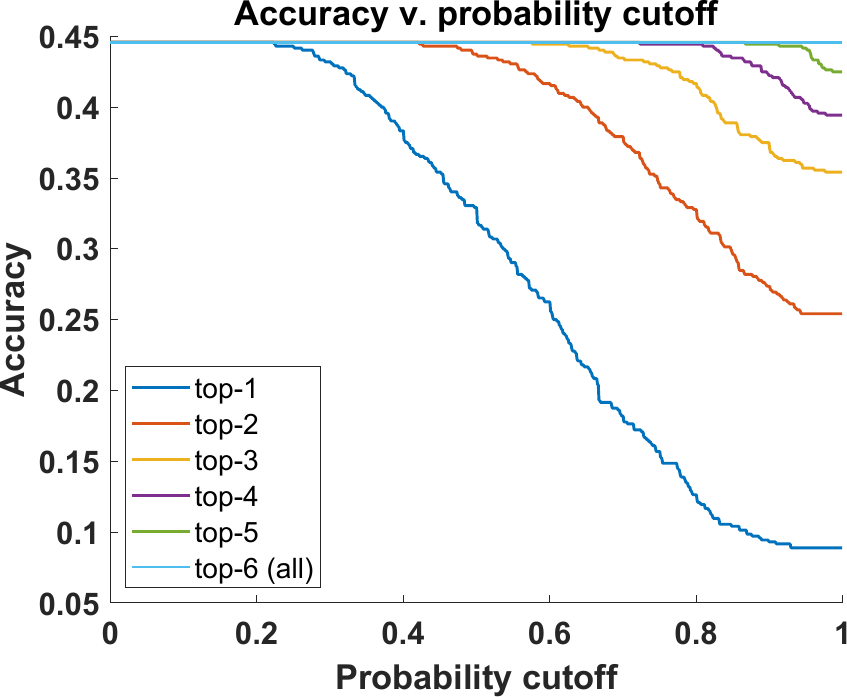}
  \caption{Top-choice accuracy as a function of different probability cutoffs for the maximum probability scenario assignments. Top-n indicates we sum the probability of the highest n scenario probability assignments. Even if the top-choice response was correct, it is marked incorrect if the probability sum is less than the cutoff. For example in the top-2 line consider an instance where the summed weight of the top 2 choices is 0.3. For probability cutoff of at least 0.3 it will be marked incorrect, regardless of whether the top choice selection was correct.}
  \label{sif:no_response}
\end{figure}

Our experiment was a forced-choice task. In Fig.~\ref*{sif:no_response} we simulate the notion that a user did not believe any response is valid. We assume that if a user assigns high probabilities to few options, they are less likely to have selected no scenario. Thus we plot the top-choice accuracy as a function of different probability cutoffs for the maximum probability scenario assignments. Top-n indicates the sum of the probability of the $n$ highest probability scenario assignments is calculated. Specifically, consider $[p(s_1), ..., p(s_6)]$ to be the sorted probability assignments for the 6 scenarios in some instance, where $p(s_1) \geq ... \geq p(s_6)$. The summed probability for that instance in the top-n case is $\sum_{i=1}^n p(s_i)$. We calculate accuracy as in Fig.~2, except even if the top-choice response was correct, it is marked incorrect if the probability sum is less than the cutoff. For example, if a user rated ``attention'' and ``gratitude'' highly, it is possible they would have selected one of those 2 rather than a ``no response'' option. The top-2 graph can be used to analyze such behaviors. By providing this analysis participants no longer need to decide at what certainty level they would select no scenario, and instead a more complete profile of user perception can be analyzed.

\subsection{First Round Classification Results}

\begin{figure}
\centering
  \includegraphics[width=\columnwidth]{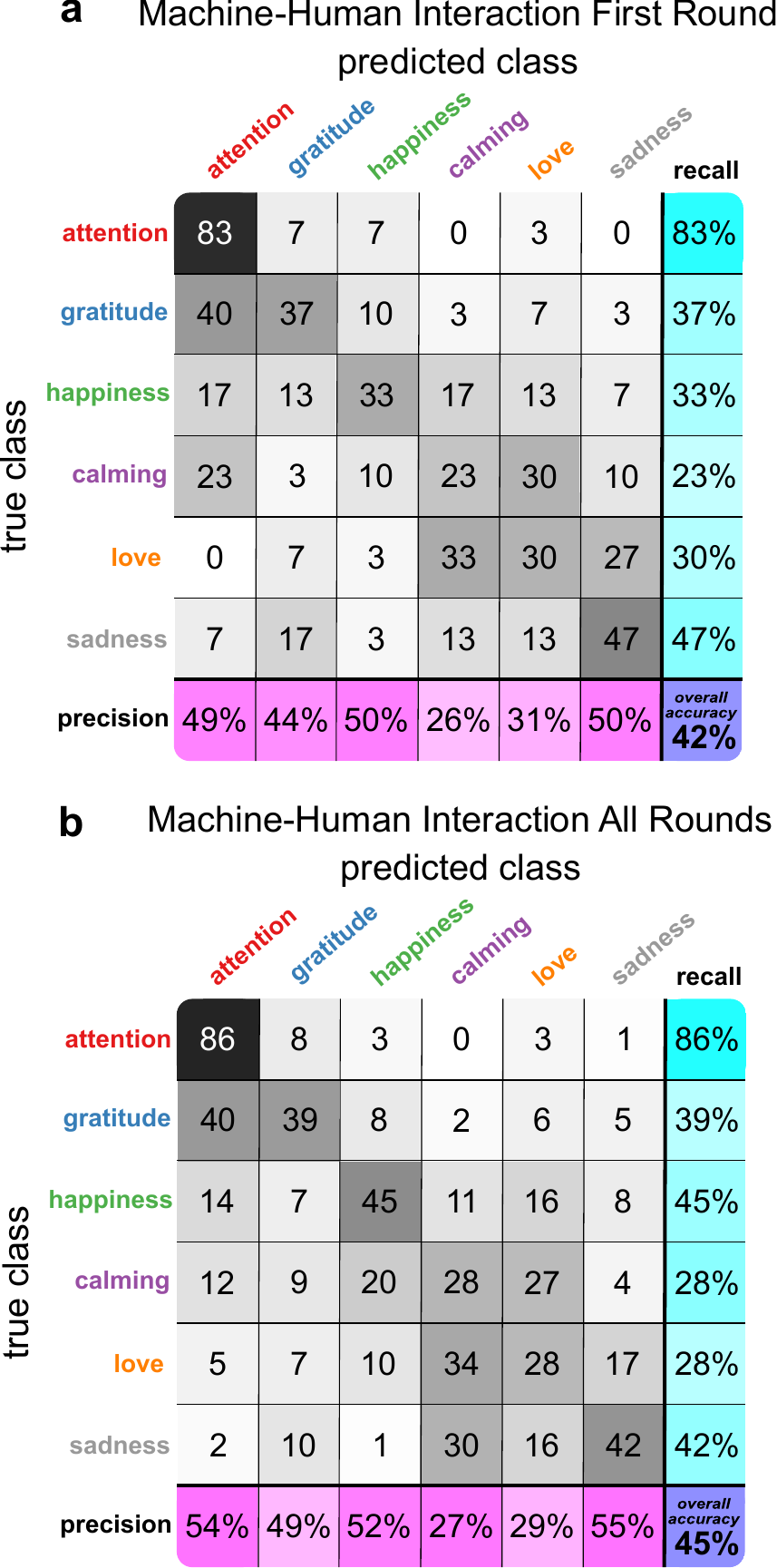}
  \caption{Comparing first round top choice classification with all rounds top choice classification. Each figure is a confusion matrix for the scenario that participants thought was most likely displayed by the rendered signal. They provide information on the participants' overall accuracy, how participants confused signals, and participants' precision and recall for their top scenario choices. (\textbf{a}) First of the four decoding rounds.  Rows are normalized to 100, with 30 samples per row in raw data. (\textbf{b}) All four decoding rounds. Rows are normalized to 100, with 120 samples per row in raw data.}
  \label{sif:first_round}
\end{figure}

For top choice classification, Fig.~\ref*{sif:first_round} compares the first round classification results with the overall result. We see 42\% accuracy for the first decoding compared to 45\% accuracy overall, and with similar areas of confusion.

\subsection{Self-Assessment Manikin clustering}

We provide clustering analysis for the Self-Assessment Manikin ratings of the displayed haptic signals (Fig.~\ref{sif:manikin_cluster}). Using the Calinski-Harabasz~\cite{SupCaliski1974ADM} criterion we find only 5 clusters, unlike the 7 obtained when clustering on the user-assigned probabilities (Fig.~8). We see two clusters similar to those obtained via the user-assigned probabilities. Cluster 4 here is similar to Cluster 7 from the user-assigned probabilities, a cluster which primarily contains points from the sadness scenario. Cluster 5 here is similar to Cluster 6 for the user-assigned probabilities, a cluster which has many points from calming, love, and sadness. While the user-assigned probabilities had clusters which seemingly exemplified each of happiness, gratitude, and attention, no such clusters exist in the Self-Assessment Manikin clustering. This may be evidence that the 2-dimensional Self-Assessment Manikin may be insufficient to fully capture common modes of interpretation by the subjects.

\subsection{Actuator Signal Measurement}

\begin{figure}
\centering
  \includegraphics[width=\columnwidth]{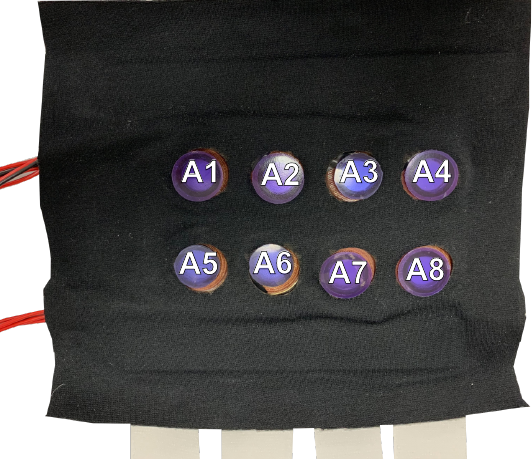}
  \caption{Voice coil actuator sleeve laid flat. Each actuator has a thin plastic covering to increase surface area. Signals are measured while sleeve is laid flat.}
  \label{sif:opensleeve}
\end{figure}

\begin{figure*}
\centering
  \includegraphics[width=\textwidth]{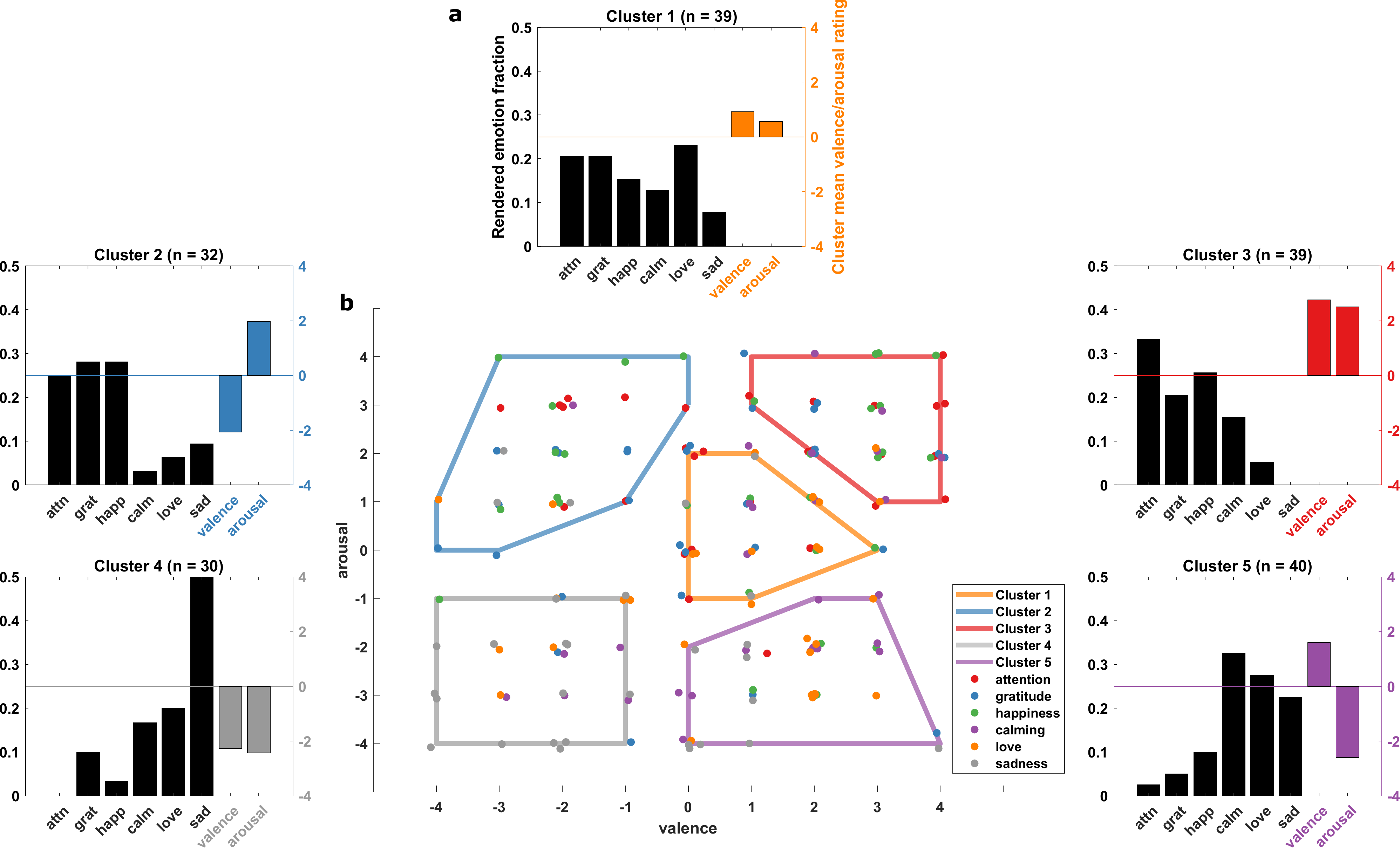}
  \caption{Self-Assessment Manikin cluster analysis. Each subject provided valence and arousal ratings for each signal, providing a 2-dimensional vector describing their score. (\textbf{a}) The outer figures show the results of a K-means analysis. The points were clustered to find consistent responses across participants. The black bars show the fraction of data points in the cluster where the signal played was from that scenario. The colored bars show the vector value of the mean of the cluster. This can be interpreted as the black bar showing the presented signal, and colored bar showing the mean valence and arousal participants indicated for signals in that cluster. (\textbf{b}) The central figure provides a cluster visualization. Each participant's valence and arousal scores are plotted in 2D. The point color represents the presented signal. Because the data is discrete, a small amount of noise is added to each point so they do not directly overlap. We then drew the convex hull of the points in each cluster from (a). This is a two-dimensional visualization of cluster locations, in order to provide intuition for what the graphs in (a) represent.}
  \label{sif:manikin_cluster}
\end{figure*}

\begin{figure*}
\centering
  \includegraphics[width=\textwidth]{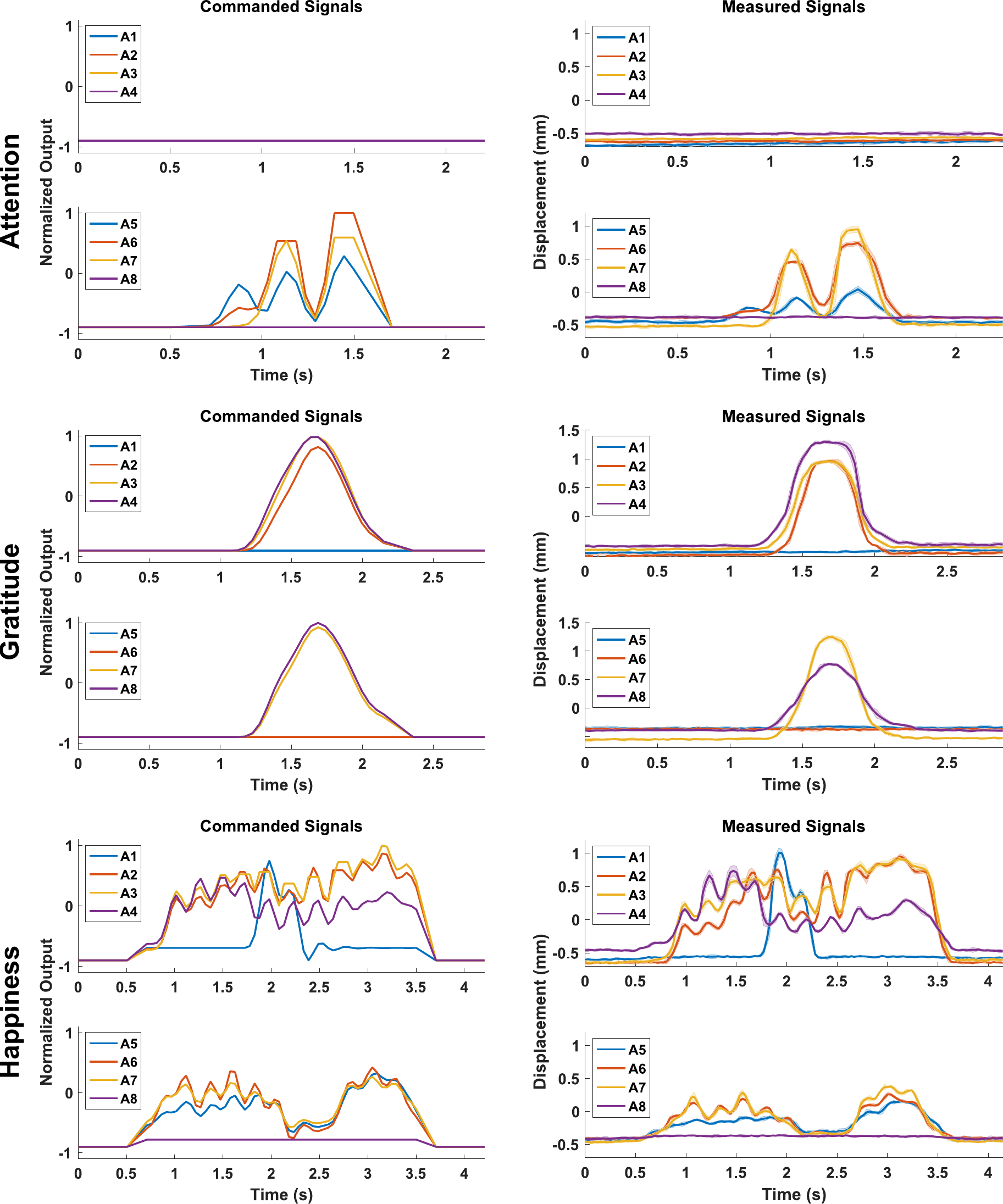}
  \caption{Commanded and measured social touch signals. The commanded (left) and measured (right) signals sent to the voice coil actuators. The shaded error bars of the measured signals represent the standard deviation of six repetitions.}
  \label{sif:signal1}
\end{figure*}

\begin{figure*}
\centering
  \includegraphics[width=\textwidth]{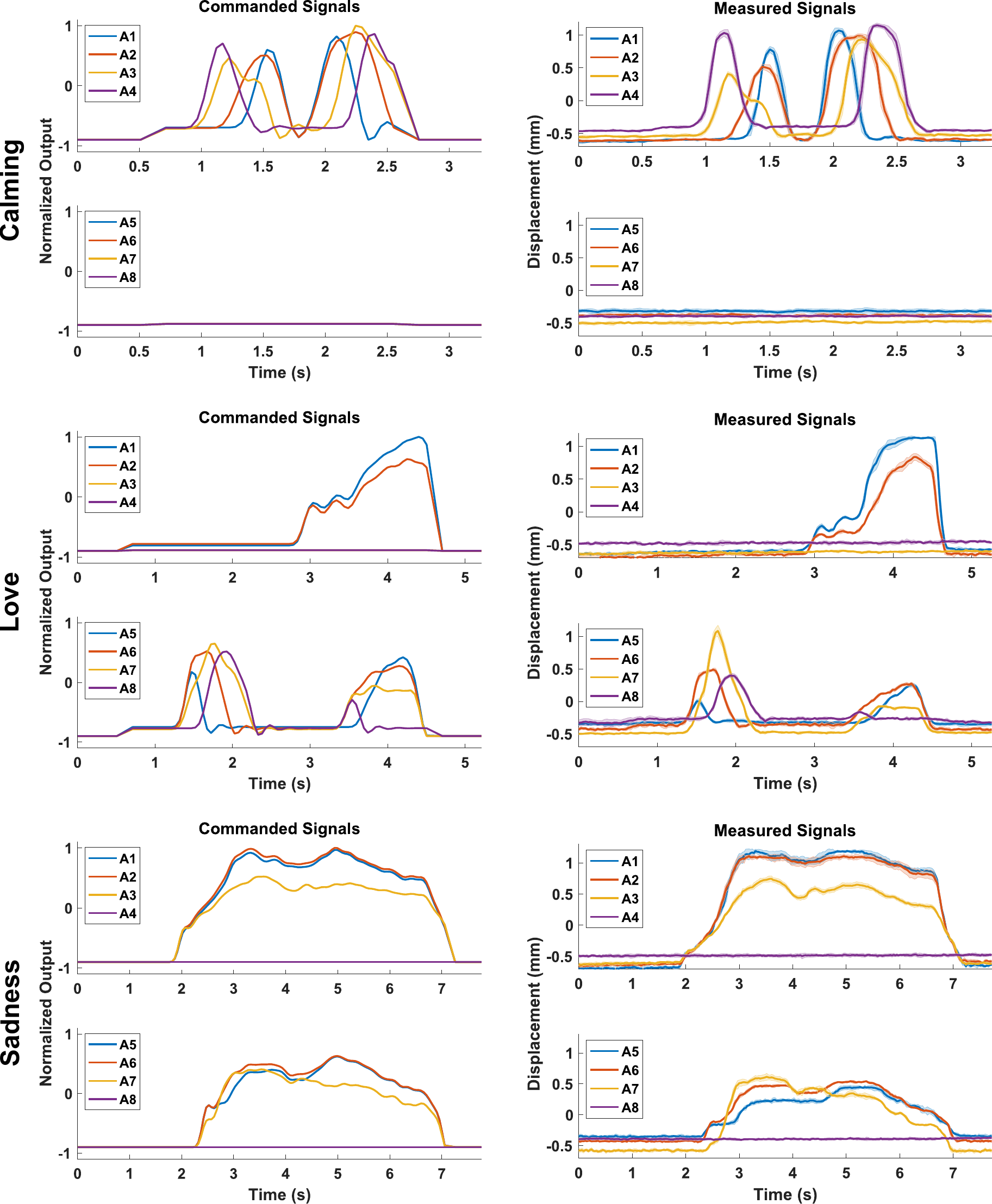}
  \caption{Commanded and measured social touch signals. The commanded (left) and measured (right) signals sent to the voice coil actuators. The shaded error bars of the measured signals represent the standard deviation of signal repetitions.}
  \label{sif:signal2}
\end{figure*}

We recorded the output of the each voice coil actuator laid flat (Fig.~\ref*{sif:opensleeve}) using the Micron Tracker Sx60 (ClaroNav, Toronto, ON, CA), a vision tracker with sub-millimeter resolution. We recorded six trials for each scenario and report the means and standard deviations of the actuator displacements (Figs.~\ref*{sif:signal1}-\ref*{sif:signal2}). We see that the actuators do not directly match the commanded signals, due to differing response characteristics for each actuator. This is possibly due to the use of voice coils that were originally designed to produce sound, where accuracy of frequency response is more important than amplitude. However, the signals sent to the actuators are repeatable across trials, as demonstrated by the small standard deviations. Integrating closed-loop control on the actuator forces or displacements using sensors is left for future work.

\clearpage
\newpage

\section{Supplementary Media, Code, and Data}

\subsection*{Movie SM1: Social Touch Recording Demonstration} This is a video showing an example interaction from our social touch dataset collection. The video shown here is a demo where two authors on the paper interact, and this data is not included in the social touch dataset. The audio is toucher version of the ``love'' scenario.

\subsection*{Movie SM2: Voice Coil Actuation} This is a video of the voice coil actuators displaying our ``love'' signal. Fig.~\ref*{sif:signal2} shows the commanded and recorded signal. We see there is a stroking motion followed by a squeezing motion.

\subsection*{Code SC1: Mapping Algorithm Code} \url{https://github.com/charm-lab/social_multiobject_tracking}

\subsection*{Dataset SD1:  Audio for the Scenario Prompts}  \url{https://stanford.box.com/v/sparse-social-touch} Folder: audio\_prompts. (text available in section SI Social Touch Dataset: Scenario Prompts).

\subsection*{Dataset SD2: Public Social Touch Dataset} We provide the recorded data for public use. The dataset consists of PPS sensor data for each touch stored as 3D Python numpy arrays. We also provide annotations for which gesture was being used during this touch. If more than one clear gesture was made, the sections with each gesture are annotated. \url{https://stanford.box.com/v/sparse-social-touch} Folder: pressure\_data.

\subsection*{Dataset SD3: Classification Experiment Results} In the scenario classification experiment we asked each participant to indicate the probability that a displayed signal was drawn from each scenario. This was repeated three times for each signal and recorded. In addition, for each signal we asked each participant to rate the valence and arousal of that signal, the results of which are available here. \url{https://stanford.box.com/v/sparse-social-touch} Folder: classification\_results. 

\bibliographystyle{IEEEtran}

\bibliography{SIBib}